\documentclass{lmcs}
\pdfoutput=1

\usepackage{lastpage}
\lmcsdoi{19}{1}{22}
\lmcsheading{}{\pageref{LastPage}}{}{}%
{Oct.~17,~2022}{Mar.~31,~2023}{}

\makeatletter
\renewcommand\paragraph{\@startsection{paragraph}{4}{\z@}%
	{2.25ex \@plus 1ex \@minus .2ex}%
	{-0.75em}%
	{\normalfont\normalsize\bfseries}}
\makeatother

\usepackage[utf8]{inputenc}
\usepackage{color}
\usepackage{xcolor}

\usepackage{eucal}
\usepackage{ifthen}
\usepackage{xargs}
\usepackage{xcolor}
\usepackage{xspace}
\usepackage{xifthen}  
\usepackage{tikz}
\usetikzlibrary{cd}
\usetikzlibrary{automata,shapes,arrows}
\usepackage{mathbbol} 
\usepackage{nicefrac}
\usepackage{hyperref}
\usepackage{thm-restate}
\usepackage{amssymb}
\usepackage{commath}
\usepackage{etoolbox} 
\usepackage{algorithm}
\usepackage[noend]{algpseudocode}

\hypersetup{
  breaklinks   = true,
  colorlinks   = true, 
  urlcolor     = blue, 
  linkcolor    = blue, 
  citecolor    = red   
}
\hypersetup{final} 

\usepackage[inline,shortlabels]{enumitem}
\newlist{inlinelist}{enumerate*}{1}
\setlist*[inlinelist,1]{%
  label=(\roman*),
}

\newcommand{\ifempty}[3]{%
  \ifthenelse{\isempty{#1}}{#2}{#3}%
}

\newenvironment{proofof}[2]{%
  \subsection*{Proof of {#1}~\ref{#2}}
  \label{#2-proof}
  }%
  {}

\newcommand{\eg}{e.g.\@\xspace}
\newcommand{\ie}{i.e.\@\xspace}
\newcommand{\wrt}{w.r.t.\@\xspace}

\newcommand{\powset}{\mathcal{P}}

\newcommand{\numN}{\mathbb{N}}
\newcommand{\numR}{\mathbb{R}}
\newcommand{\setenum}[1]{\{#1\}}
\newcommand{\setcomp}[2]{\{ {#1} \,\mid\, {#2}\}}

\newcommand{\labels}{\mathcal{L}}
\newcommand{\labelL}{l}
\newcommand{\lab}{\ell} 
\newcommand{\tsQ}{\mathcal{Q}}

\newcommand{\stepsN}{n}
\newcommand{\stepsNi}{n'}
\newcommand{\stepsNb}{\bar{n}}
\newcommand{\distD}{\delta}
\newcommand{\distDi}{\delta'}
\newcommand{\distDii}{\delta''}
\newcommand{\boundK}{k}
\newcommand{\crysim}[2]{\sim^{#1}_{#2}}
\newcommand{\cryset}[3]{\sim^{#1}_{#2}(#3)}

\newcommand{\stateP}{p}
\newcommand{\stateQ}{q}
\newcommand{\stateQi}{q'}
\newcommand{\stateQii}{q''}
\newcommand{\sP}{\stateP}
\newcommand{\sPi}{p'}
\newcommand{\sPii}{p''}
\newcommand{\sQ}{\stateQ}
\newcommand{\sQi}{\stateQi}
\newcommand{\statesQ}{Q}
\newcommand{\tsPrName}{\Pr}
\newcommand{\tsPr}[2]{\tsPrName(#1,#2)}
\newcommand{\tracesT}{T}
\newcommand{\traceT}{t}

\newcommand{\finTraceT}{\tilde{t}}

\newcommand{\finTraceU}{\tilde{u}}

\newcommand{\finTracesT}{\tilde{T}}
\newcommand{\cyl}[1]{\mathit{Cyl}({#1})}
\newcommand{\trStart}[1]{\cyl{#1}}
\newcommand{\fram}[2]{F^{#2}_{#1}}

\newcommand{\flow}{\mathit{flow}}

\newcommand{\sSetP}{P}
\newcommand{\card}[1]{|#1|}

\newcommand{\CSim}[2]{\crysim{#1}{#2}}
\newcommand{\R}[3]{\cryset{#1}{#2}{#3}}
\newcommand{\TR}[3]{\ifempty{#1}{\tilde{R}}{\ifempty{#3}{\tilde{R}_{#2}^{#1}}{\tilde{R}_{#2}^{#1}({#3})}}}
\newcommand{\NNR}{\mathbb{R}_0^+}
\newcommand{\intervalCC}[2]{[#1,#2]}
\renewcommand{\prob}[2]{\ifempty{#1} {\Pr} {\ifempty{#2}{\Pr(#1)}{\Pr(#1,#2)}}}

\newcommand{\logTrue}{\sf true}
\newcommand{\logNext}{{\sf X}}
\newcommand{\logUntil}{{\sf U}}
\newcommand{\logPhi}{\phi}
\newcommand{\logPhii}{\phi'}
\newcommand{\logPsi}{\psi}

\newcommand{\logPr}[2]{{\sf Pr}_{#1}[#2]}
\newcommand{\probP}{\pi}
\newcommand{\dirR}{r}
\newcommand{\sem}[4]{[\![ {#4} ]\!]^{#1}_{#2, #3}}

\newcommand{\nestMax}[2]{{#1}_{\it max}{(#2)}}

\newcommand{\famF}{\Xi}
\newcommand{\simFam}{\equiv}

\newcommand{\sat}[3]{\models_{\ifempty{#3}{#1}{{#1},{#3}}}^{#2}}
\newcommand{\lUntil}{\logUntil}
\newcommand{\lNext}{\logNext}

\newcommand{\sFormula}[1][]{\logPhi_{#1}}
\newcommand{\sFormulai}[1][]{\logPhii_{#1}}
\newcommand{\pFormula}[1][]{\logPsi_{#1}}
\newcommand{\lTrue}{\logTrue}
\newcommand{\lNot}{\mathtt{\neg}}
\newcommand{\lAnd}{\mathtt{\land}}
\newcommand{\atomA}[1][]{a_{#1}}

\newtheorem{applemma}{Lemma}[section]{\bfseries}{\itshape}


\begin{document}

\author[M.~Bartoletti]{Massimo Bartoletti\lmcsorcid{0000-0003-3796-9774}}[a]

\author[M. Murgia]{Maurizio Murgia\lmcsorcid{0000-0001-7613-621X}}[b]

\author[R. Zunino]{Roberto Zunino\lmcsorcid{0000-0002-9630-429X}}[c]

\address{University of Cagliari, Cagliari, Italy}
\email{bart@unica.it}

\address{Gran Sasso Science Institute, L'Aquila, Italy}
\email{maurizio.murgia@gssi.it}

\address{Universit\`a degli Studi di Trento, Trento, Italy}
\email{roberto.zunino@unitn.it}

\title[Probabilistic bisimulations for PCTL]{Sound approximate and asymptotic \texorpdfstring{\\}{} probabilistic bisimulations for PCTL}

\maketitle

\begin{abstract}
  We tackle the problem of establishing the soundness of approximate
  bisimilarity with respect to PCTL and its relaxed semantics.
  To this purpose, we consider a notion of bisimilarity inspired by
  the one introduced by Desharnais, Laviolette, and Tracol, and
  parametric with respect to an approximation error $\distD$, and to
  the depth $\stepsN$ of the observation along traces.
  Essentially, our soundness theorem establishes that, when a state
  $\stateQ$ satisfies a given formula up-to error $\distD$ and steps
  $\stepsN$, and $\stateQ$ is bisimilar to $\stateQi$ up-to error
  $\distDi$ and enough steps, we prove that $\stateQi$ also satisfies
  the formula up-to a suitable error $\distDii$ and steps $\stepsN$.
  The new error $\distDii$ is computed from $\distD,\distDi$
  and the formula, and only depends linearly on $\stepsN$.
  We provide a detailed overview of our soundness proof.

  We extend our bisimilarity notion to families of states, thus
  obtaining an asymptotic equivalence on such families.
  We then consider an asymptotic satisfaction relation for PCTL
  formulae, and prove that asymptotically equivalent families of
  states asymptotically satisfy the same formulae.
\end{abstract}

\section{Introduction}

The behaviour of many real-world systems can be formally modelled as probabilistic processes, \eg as discrete-time Markov chains.
Specifying and verifying properties on these systems requires probabilistic versions of temporal logics, such as PCTL~\cite{HanssonJonsson94}.
PCTL allows to express probability bounds using the formula $\logPr{\geq \probP}{\logPsi}$, which is satisfied by those states starting from which the path formula $\logPsi$ holds with probability $\geq \probP$.
A well-known issue is that real-world systems can have tiny deviations from their mathematical models, while logical properties, such as those written in PCTL, impose sharp constraints on the behaviour.
To address this issue, one can use a \emph{relaxed} semantics for PCTL, as in~\cite{DInnocenzo12hscc}. 
There, the semantics of formulae is parameterised over the error 
$\distD\geq 0$ one is willing to tolerate. 
While in the standard semantics of $\logPr{\geq \probP}{\logPsi}$
the bound $\geq\probP$ is \emph{exact}, 
in relaxed PCTL this bound is weakened to $\geq\probP-\distD$.
So, the relaxed semantics generalises the standard PCTL semantics of~\cite{HanssonJonsson94}, which can be obtained by choosing $\distD=0$.
Instead, choosing an error $\distD > 0$ effectively provides a way to measure ``how much'' a state satisfies a given formula: some states might require only a very small error, while others a much larger one.

When dealing with temporal logics such as PCTL, one often wants to study some notion of state equivalence which preserves the semantics of formulae: that is, when two states are equivalent, they satisfy the same formulae.
For instance, probabilistic bisimilarities like those in~\cite{Desharnais10iandc,Desharnais02iandc,Larsen91iandc} preserve the semantics of formulae for PCTL and other temporal logics.
Although \emph{strict} probabilistic bisimilarity preserves 
the semantics of relaxed PCTL, 
it is not \emph{robust} against small deviations in 
the probability of transitions in Markov chains~\cite{Giacalone90ifip2}.
A possible approach to deal with this issue is to also
relax the notion of probabilistic bisimilarity, 
by making it parametric with respect to an error $\distD$~\cite{DInnocenzo12hscc}.
Relaxing bisimilarity in this way poses a choice regarding
which properties of the strict probabilistic bisimilarity are to be kept. 
In particular, transitivity is enjoyed by the strict probabilistic bisimilarity,
but it is \emph{not} desirable for the relaxed notion.
Indeed, we could have three states $\stateQ,\stateQi$ and $\stateQii$ where the behaviour of $\stateQ$ and $\stateQi$ is similar enough (within the error $\distD$), the behaviour of $\stateQi$ and $\stateQii$ is also similar enough (within $\distD$), but the distance between $\stateQ$ and $\stateQii$ is larger than the allowed error $\distD$.
At best, we can have a sort of ``triangular inequality'', where $\stateQ$ and $\stateQii$ can still be related but only with a larger error $2\cdot\distD$.

\begin{figure}[t]
  \centering
  \begin{tikzpicture}[scale=0.95, transform shape, >=latex]
    \node [state] (H) at (0, 0) {$\mathit{head}$};
    \path [->] (H) edge [loop above] node {\scriptsize $\nicefrac{1}{2}$} (H);
    \node [state] (T) at (5, 0) {$\mathit{tail}$};
    \path [->] (T) edge [loop above] node {\scriptsize $\nicefrac{1}{2}$} (T);
    \path [->] (T.south west) edge [out=195, in=-15] node[below] {$\nicefrac{1}{2}$} (H.south east);
    \path [->] (H.north east) edge [out=15, in=165] node[above] {$\nicefrac{1}{2}$} (T.north west);
  \end{tikzpicture}
  \caption{A Markov chain modelling repeated tosses of a fair coin.} 
  \label{fig:fair-coin}
\end{figure}
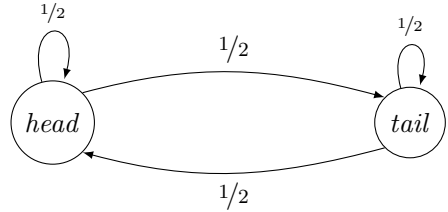

Bisimilarity is usually defined by coinduction, essentially requiring that the relation is preserved along an arbitrarily long sequence of moves.
Still, in some settings, observing the behaviour over a very long run 
is undesirable.
For instance, consider the PCTL formula 
$\logPhi = \logPr{\geq 0.5}{{\sf true}\ \logUntil^{\leq \stepsN}\ {\sf a}}$,
which is satisfied by those states from which, with probability $\geq 0.5$,
${\sf a}$ is satisfied within $\stepsN$ steps.
In this case, a behavioural equivalence relation 
that preserves the semantics of $\logPhi$ can neglect the long-run behaviour 
after $n$ steps.
More generally, if all the until operators are \emph{bounded}, 
as in $\logPhi_1 \logUntil^{\leq k} \logPhi_2$,
then each formula has an upper bound of steps $\stepsN$ after which
a behavioural equivalence relation can ignore what happens next.
Observing the behaviour after this upper bound is unnecessarily strict,
and indeed in some settings it is customary to neglect 
what happens in the very long run.
For instance, a real-world player repeatedly tossing a coin is usually 
considered equivalent to a Markov chain with two states 
and four transitions with probability $\nicefrac{1}{2}$
(see~\autoref{fig:fair-coin}), 
even if in the long run the real-world system will diverge from the ideal one
(\eg, when the player dies).

Another setting where observing the long-term behaviour is notoriously 
undesirable is that of cryptography.
When studying the security of systems modelling cryptographic protocols, 
two states are commonly considered equivalent when their behaviour is similar 
(up to a small error $\distD$) in the short run, 
even when in the very long run they diverge.
For instance, a state $\stateQ$ could represent an ideal system where no attacks can be performed by construction, while another state $\stateQi$ could represent a real system where an adversary can try to disrupt the cryptographic protocol.
In such a scenario, if the protocol is secure, we would like to have $\stateQ$ and $\stateQi$ equivalent, since the behaviour of the real system is close to the one of the ideal system.
Note that in the real system an adversary can repeatedly try to guess the secret cryptographic keys, and break security in the very long run, with very high probability.
Accordingly, standard security definitions require that the behaviour of the ideal and real system are within a small error, but only for a \emph{bounded} number of steps, after which their behaviour could diverge.

\paragraph{Contributions}

To overcome the above mentioned issues, 
in this work we introduce a bounded, approximate notion of bisimilarity
$\crysim{\stepsN}{\distD}$, that only observes the first $\stepsN$ steps,
and allows for an error $\distD$.
Unlike standard bisimilarity, our relation is naturally defined by \emph{induction} 
on $\stepsN$.
We call this looser variant of bisimilarity an \emph{up-to-$\stepsN,\distD$} bisimilarity.
We showcase up-to-$\stepsN,\distD$ bisimilarity on a running example
(Examples~\ref{ex:pctl:padlock}, \ref{ex:sim:padlock}, \ref{ex:results:padlock}, and~\ref{ex:asymptotic:padlock}),
comparing an ideal combination padlock against a real one which can be opened by
an adversary guessing its combination.
We show that the two systems are bisimilar up-to-$\stepsN,\distD$,
while they are not bisimilar according to the standard coinductive notion.
We then discuss how the two systems satisfy a
basic security property expressed in PCTL, with suitable errors.
To make our theory amenable to reason about infinite-state systems,
such as those usually found when modelling cryptographic protocols,
all our results apply to Markov chains with countably many states.
In this respect, our work departs from most literature on 
probabilistic bisimulations~\cite{DInnocenzo12hscc,Song13lmcs} 
and bisimilarity distances~\cite{Breugel17siglog,TangB17concur,TangB18cav,TangB16concur,Fu12icalp,ChenBW12fossacs,BreugelSW08lmcs},
which usually assume \emph{finite}-state Markov chains,
as they focus on computing the distances.
In~\autoref{ex:sim:pingpong} we exploit infinite-state Markov chains to 
compare a biased random bit generator with an ideal one.

Our first main contribution is a soundness theorem establishing that,
when a state $\stateQ$ satisfies a PCTL formula $\logPhi$ (up to a
given error), any bisimilar state $\stateQi \crysim{}{} \stateQ$ must
also satisfy $\logPhi$, at the cost of a slight increase of the error.
More precisely, if $\logPhi$ only involves until operators
bounded by $\leq\stepsN$, state $\stateQ$ satisfies $\logPhi$ up to some
error, and bisimilarity holds for enough steps and error $\distD$,
then $\stateQi$ satisfies $\logPhi$ with an \emph{additional}
asymptotic error $O(\stepsN\cdot\distD)$.

This asymptotic behaviour is compatible with the usual assumptions of computational security in cryptography.
There, models of security protocols include a security parameter $\eta$, which affects the length of the cryptographic keys and the running time of the protocol:
more precisely, a protocol is assumed to run for $\stepsN(\eta)$ steps, which is polynomially bounded \wrt $\eta$.
As already mentioned above, cryptographic notions of security do not
observe the behaviour of the systems after this bound $\stepsN(\eta)$,
since in the long run an adversary can surely guess the secret keys by
brute force.
Coherently, a protocol is considered to be secure if (roughly) its
actual behaviour is \emph{approximately} equivalent to the ideal one
for $\stepsN(\eta)$ steps and up to an error $\distD(\eta)$, which has
to be a negligible function, asymptotically approaching zero faster
than any rational function.
Under these bounds on $\stepsN$ and $\distD$, the asymptotic error $O(\stepsN\cdot\distD)$ in our soundness theorem is negligible in $\eta$.
Consequently, if two states $\stateQ$ and $\stateQi$ represent the ideal and actual behaviour, respectively, and they are bisimilar up to a negligible error, they will satisfy the same PCTL formulae with a negligible error.

We formalise this reasoning by providing a notion of \emph{asymptotic
  equivalence}.
We start by considering families of states $\famF(\eta)$, intuitively
representing the behaviour of a system depending on a security
parameter $\eta$.
Our asymptotic equivalence $\famF_1 \simFam \famF_2$ holds whenever
the behaviour of the two families is $\stepsN,\distD$-bisimilar within
a negligible error whenever we only perform a polynomial number of
steps.
We further introduce an \emph{asymptotic satisfaction relation}
$\famF \models \logPhi$ which holds whenever the state $\famF(\eta)$
satisfies $\logPhi$ under similar assumptions on the number of steps
and the allowed error.
Our second main result is the soundness of the asymptotic equivalence
with respect to asymptotic satisfaction.
Asymptotically equivalent families asymptotically satisfy the same
PCTL formulae.

We provide a detailed overview of the proof of our soundness theorem
for $\stepsN,\distD$-bisimilarity in~\autoref{sec:result},
deferring the gory technicalities to~\autoref{sec:proofs}.
The proof of asymptotic soundness, which exploits the
soundness theorem for $\stepsN,\distD$-bisimilarity, is given
in~\autoref{sec:asymptotic}.

\section{Related work}

There is a well-established line of research on 
establishing soundness and completeness of probabilistic bisimulations 
against various kinds of probabilistic logics
\cite{Desharnais10iandc,FurberMM19lmcs,Hermanns11iandc,Larsen91iandc,Mio17fuin,Mio18lics}.

The work closest to ours is that of D’Innocenzo, Abate and
Katoen~\cite{DInnocenzo12hscc}, which addresses the model checking
problem on a relaxed PCTL differing from ours in a few aspects.
First, their syntax allows for an individual bound
on the number of steps $k$
for each until operator $\logUntil^{\leq k}$, while we assume all such
bounds are equal
and we make the semantics of PCTL parametrized \wrt the number of steps to be considered in until.
This approach allows us to simplify the statement of the soundness theorem and the definition of asymptotic satisfaction relation,
since the bound is not fixed by the formula, but it is a parameter of the semantics. Dealing with the case where each until in a formula could have its bound seems possible, at the cost of increasing the level of technicalities.
Second, their main result shows that bisimilar states
up-to a given error $\epsilon$
satisfy the same formulae $\logPsi$, provided that $\logPsi$ ranges
over the so-called $\epsilon$-robust formulae.  Instead, our soundness
result applies to \emph{all} PCTL formulae, and ensures that when
moving from a state satisfying $\logPhi$ to a bisimilar one, $\logPhi$
is still satisfied, but at the cost of slightly increasing the error.
Third, their relaxed semantics differs from ours. In ours, we relax
all the probability bounds by the same amount $\distD$. Instead, the
relaxation in~\cite{DInnocenzo12hscc} affects the bounds by a
different amount which depends on the error~$\epsilon$, the until
bound $k$, and the underlying DTMC.

Desharnais, Laviolette and Tracol~\cite{Desharnais08qest} use a
coinductive approximate probabilistic bisimilarity, up-to an error
$\distD$.
Using such coinductive bisimilarity, \cite{Desharnais08qest}
establishes the soundness and completeness with respect to a
Larsen-Skou logic~\cite{Larsen91iandc} (instead of PCTL).
In~\cite{Desharnais08qest}, a bounded, up-to $\stepsN,\distD$ version
of bisimilarity is only briefly used to derive a decision algorithm
for coinductive bisimilarity under the assumption that the state space
is finite.
In our work, instead, the bounded up-to $\stepsN,\distD$ bisimilarity
is the main focus of study.
In particular, our soundness result only assumes $\stepsN,\distD$
bisimilarity, which is strictly weaker than coinductive bisimilarity.
Another minor difference is that~\cite{Desharnais08qest} considers a
labelled Markov process, \ie the probabilistic variant of a labelled
transition system, while we instead focus on DTMCs having labels on
states.

Bian and Abate~\cite{Bian17fossacs} study bisimulation and trace
equivalence up-to an error $\epsilon$, and show that
$\epsilon$-bisimilar states are also $\epsilon’$-trace equivalent for
a suitable $\epsilon’$ which depends on~$\epsilon$.
Furthermore, they
show that $\epsilon$-trace equivalent states satisfy the same formulae
in a bounded LTL, up-to a certain error.  In our work, we focus
instead on the branching logic PCTL.

A related research line is that on \emph{bisimulation metrics}
\cite{Breugel17siglog,BreugelHMW05icalp,BreugelW05tcs}.
Some of these metrics, like our up-to bisimilarity, take approximations into
account~\cite{Desharnais99concur,Castiglioni16qapl}.  
Similarly to our
bisimilarity, bisimulation metrics allow to establish two states
equivalent up-to a certain error (but usually do not take into account
the bound on the number of steps).  Interestingly, Castiglioni, Gebler
and Tini~\cite{Castiglioni16qapl} introduce a notion of distance
between Larsen-Skou formulae, and prove that the bisimulation distance
between two processes corresponds to the distance between their
mimicking formulae.  
De Alfaro, Majumdar, Raman and Stoelinga~\cite{deAlfaroMRS08} 
elegantly characterise bisimulation metrics with a quantitative 
$\mu$-calculus. 
Such logic allows to specify interesting properties such as maximal reachability and safety probability, and the maximal probability of satisfying a general 
$\omega$-regular specification, but not full PCTL.
Mio~\cite{Mio14fossacs} characterises a bisimulation metric based on total variability
with a more general quantitative $\mu$-calculus, 
dubbed {\L}ukasiewicz 
$\mu$-calculus, able to encode PCTL.
Both \cite{deAlfaroMRS08} and~\cite{Mio14fossacs} do not take 
the number of steps into account,
therefore their applicability to the analysis of security protocols 
is yet to be investigated.

Metrics with discount~\cite{Desharnais04tcs,deAlfaro03icalp,Bacci21lmcs,DengCPP06entcs,BreugelSW08lmcs} 
are sometimes used to relate the behaviour of probabilistic processes, 
weighing less those events that happen in the far future compared to those 
happening in the first steps.
Often, in these metrics each step causes the probability of the next events to be multiplied by a constant factor $c < 1$, in order to diminish their importance. Note that this discount makes it so that after $\eta$ steps, this diminishing factor becomes $c^\eta$, which is a negligible function of $\eta$. As discussed before, in cryptographic security one needs to consider as important those events happening within polynomially many steps, while neglecting the ones after such a polynomial threshold. Using an exponential discount factor $c^\eta$ after only $\eta$ steps goes against this principle, since it would cause a secure system to be at a negligible distance from an insecure one which can be violated after just $\eta$ steps.
For this reason, instead of using a metric with discount, in this paper we resort to a bisimilarity that is parametrized over the number of steps $\stepsN$ and error $\distD$, allowing us to obtain a notion which distinguishes between the mentioned secure and insecure systems.

Several works develop algorithms to decide probabilistic bisimilarity,
and to compute metrics
\cite{BreugelW14birthday,ChenBW12fossacs,Fu12icalp,TangB16concur,TangB17concur,TangB18cav}.
To this purpose, they restrict to finite-state systems,
like \eg probabilistic automata.
Our results, instead, apply also to infinite-state systems.

In \cite{ZuninoD05} a calculus with cryptographic primitives is
introduced, together with a semantics where attackers have a
probability $\probP(\eta)$ of guessing encryption keys.
It is shown that, assuming that $\probP(\eta)$ is negligible and that attackers
run in polynomial time, some security properties (\eg secrecy,
authentication) are equivalent to the analogous properties with
standard Dolev-Yao assumptions (that is, attackers never guess keys
but are not restricted to polynomial time).
This result can be seen as a special case of our asymptotic soundness
theorem.

The interesting work \cite{LagoG22} proposes a behavioural notion of
indistinguishability between session typed probabilistic
$\pi$-calculus processes, with the aim of providing a formal system
for proving security of real cryptographic protocols by comparison
with ideal ones. The type system, which is based on bounded linear
logic \cite{GirardSS92,LagoG16}, guarantees that processes terminate
in polynomial time.
This differs from our approach, where polynomiality appears directly
in the equivalence definition (\autoref{def:crysim}).
Moreover, the calculus of
\cite{LagoG22} is quite restrictive: for instance, it is not possible to
specify adversaries that access an oracle a polynomial number of times.
By contrast, our abstract model is general enough to represent such
adversaries.

\paragraph{Comparison with~\cite{BMZ22coordination}}

This paper extends the work~\cite{BMZ22coordination} in two
directions.
First, the current paper includes the proofs of all statements,
which were not present in~\cite{BMZ22coordination}.
Second, in~\cite{BMZ22coordination} we hinted at the possible
application of soundness to the asymptotic behaviour of systems which
depend on a parameter $\eta$.
Here, we properly develop and formalise that intuition 
in~\autoref{sec:asymptotic}, providing a new asymptotic soundness result.

\section{The probabilistic temporal logic PCTL}

Assume a set $\labels$ of labels, ranged over by $\labelL$,
and let $\distD,\probP$ range over non-negative reals.
A \emph{discrete-time Markov chain} (DTMC) is a standard model of probabilistic
systems. Throughout this paper, we consider a DTMC having a countable,
possibly infinite, set of states $\stateQ$, each carrying a subset
of labels $\lab(\stateQ) \subseteq \labels$.

\begin{defi}[Discrete-Time Markov Chain]
  \label{def:pctl:dtmc}
  A (labelled) DTMC is a triple $(\tsQ, \tsPrName, \lab)$ where:
  \begin{itemize}
  \item $\tsQ$ is a countable set of states;
  \item $\tsPrName : \tsQ^2 \to [0,1]$ is a function, named transition
    probability function;
  \item $\lab : \tsQ \to \powset(\labels)$ is a labelling function
  \end{itemize}
  Given $\stateQ \in \tsQ$ and $\statesQ \subseteq \tsQ$, 
  we write $\tsPr{\stateQ}{\statesQ}$ for
  $\sum_{\stateQi\in\statesQ} \tsPr{\stateQ}{\stateQi}$
  and we require that $\tsPr{\stateQ}{\tsQ}=1$ for all $\stateQ\in\tsQ$.
\end{defi}

A \emph{trace} is an infinite sequence of states
$\traceT = \stateQ_0\stateQ_1\cdots$,
where we write $\traceT(i)$ for $\stateQ_i$,
\ie the $i$-th element of $\traceT$.
A \emph{trace fragment} is a finite, non-empty sequence of
states $\finTraceT = \stateQ_0 \cdots \stateQ_{n-1}$, 
where $\card{\finTraceT}= n\geq 1$ is its length.
Given a trace fragment $\finTraceT$ and a state $\stateQ$, we write
$\finTraceT\stateQ^\omega$ for the trace
$\finTraceT\stateQ\stateQ\stateQ\cdots$.

It is well-known that, given an initial state $\stateQ_0$, the DTMC
induces a $\sigma$-algebra of measurable sets of traces $\tracesT$
starting from $\stateQ_0$, \ie~the $\sigma$-algebra generated by
cylinder sets~\cite{BaierKatoen08}.
More in detail, given a trace fragment
$\finTraceT = \stateQ_0 \cdots \stateQ_{n-1}$, its \emph{cylinder set}
\[
\cyl{\finTraceT}
\; = \;
\setcomp{\traceT}{\text{$\finTraceT$ is a prefix of $\traceT$}}
\]
is given probability:
\[
\Pr(\cyl{\finTraceT})
\; = \;
\prod_{i=0}^{n-2}   \tsPr{\stateQ_i}{\stateQ_{i+1}}
\]
As usual, if $n=1$ the product is empty and evaluates to $1$.
Closing the family of cylinder sets under countable unions and complement
we obtain the family of measurable sets.
The probability measure on cylinder sets then uniquely extends to all the
measurable sets.

Given a set of trace fragments $\finTracesT$, all starting from the
same state $\stateQ_0$ and having the same length, we let
\(
  \Pr(\finTracesT) =
  \Pr(\bigcup_{\finTraceT\in\finTracesT}
  \cyl{\finTraceT})
  = \sum_{\finTraceT\in\finTracesT} \Pr(\cyl{\finTraceT})
\).
Note that using same-length trace fragments ensures that their
cylinder sets are disjoint, hence the second equality holds.

Below, we define PCTL formulae. Our syntax is mostly standard,
except for the \emph{until} operator.
There, for the sake of simplicity, we do not bound the number of steps
in the syntax $\logPhi_1\ \logUntil\ \logPhi_2$, but we do so in the
semantics.
Concretely, this amounts to imposing the same bound to
\emph{all} the occurrences of $\logUntil$ in the formula.
Such bound is then provided as a parameter to the semantics.

\begin{defi}[PCTL Syntax]
  The syntax of PCTL is given by the following grammar, defining
  \emph{state formulae} $\logPhi$ and \emph{path formulae} $\logPsi$:
  \begin{align*}
    \logPhi 
    & ::=
      \labelL
      \mid {\sf true}
      \mid \lnot \logPhi
      \mid \logPhi \land \logPhi
      \mid \logPr{\rhd \probP}{\logPsi}
      \qquad \mbox{ where } \rhd \in \setenum{>,\geq}
    \\
    \logPsi 
    & ::=
      \logNext\ \logPhi
      \mid \logPhi\ \logUntil\ \logPhi
  \end{align*}
  As syntactic sugar, we write $\logPr{< \probP}{\logPsi}$ for
  $\lnot\logPr{\geq \probP}{\logPsi}$, and $\logPr{\leq \probP}{\logPsi}$
  for $\lnot\logPr{> \probP}{\logPsi}$.
\end{defi}

Given a PCTL formula $\logPhi$, we define its 
maximum $\logNext$-nesting $\nestMax{\logNext}{\logPhi}$
and its maximum $\logUntil$-nesting $\nestMax{\logUntil}{\logPhi}$ 
inductively as follows:
\begin{defi}[Maximum Nesting]
  For $\circ \in \setenum{\logNext,\logUntil}$, we define:
  \[
  \begin{array}{c}
    \nestMax{\circ}{\labelL}
    = 
    0
    \qquad
    \nestMax{\circ}{\logTrue} 
    = 
    0
    \qquad
    \nestMax{\circ}{\lnot\logPhi}
    =
    \nestMax{\circ}{\logPhi}
    \\[8pt]
    \nestMax{\circ}{\logPhi_1 \land \logPhi_2}
    =
    \max(\nestMax{\circ}{\logPhi_1},\nestMax{\circ}{\logPhi_2})
    \qquad
    \nestMax{\circ}{\logPr{\rhd \probP}{\logPsi}}
    = 
    \nestMax{\circ}{\logPsi}
    \\[8pt]
    \nestMax{\circ}{\logNext \logPhi}
    = 
    \nestMax{\circ}{\logPhi} + 
    \begin{cases}
      1 & \text{if $\circ = \logNext$} \\
      0 & \text{otherwise}
    \end{cases}
    \\[16pt]
    \nestMax{\circ}{\logPhi_1 \logUntil \logPhi_2}
    = 
    \max(\nestMax{\circ}{\logPhi_1},\nestMax{\circ}{\logPhi_2}) +
    \begin{cases}
      1 & \text{if $\circ = \logUntil$} \\      
      0 & \text{otherwise}
    \end{cases}
  \end{array}
  \]
\end{defi}

We now define a semantics for PCTL where the probability bounds
$\rhd \probP$ in $\logPr{\rhd \probP}{\logPsi}$ can be relaxed or
strengthened by an error $\distD$.
Our semantics is parameterized over the \emph{until} bound $\stepsN$,
the error $\distD\in\numR^{\geq 0}$, and a direction
$\dirR\in\setenum{+1,-1}$.
Given the parameters, the semantics associates each PCTL state formula
with the set of states satisfying it.
Intuitively, when $\dirR = +1$ we relax the semantics of the formula,
so that increasing $\distD$ causes more states to satisfy it.  More
precisely, the probability bounds $\rhd\probP$ in positive occurrences
of $\logPr{\rhd \probP}{\logPsi}$ are decreased by $\distD$, while
those in negative occurrences are increased by $\distD$.
Dually, when $\dirR = -1$ we strengthen the semantics, modifying
$\rhd\probP$ in the opposite direction.
Our semantics is inspired by the relaxed / strengthened PCTL semantics
of~\cite{DInnocenzo12hscc}.

\begin{defi}[PCTL Semantics]
  \label{def:pctl:sem}
  The semantics of PCTL formulae is given below. Let
  $\stepsN \in \numN$, $\distD\in\numR^{\geq 0}$
  and $\dirR \in \setenum{+1,-1}$.
  \[
  \begin{array}{ll}
    \sem{\stepsN}{\distD}{\dirR}{\labelL}
    &= \setcomp{\stateQ\in\tsQ}{\labelL\in\lab(\stateQ)}
    \\
    \sem{\stepsN}{\distD}{\dirR}{\sf true}
    &= \tsQ
    \\
    \sem{\stepsN}{\distD}{\dirR}{\lnot\logPhi}
    &=
    \tsQ \setminus \sem{\stepsN}{\distD}{-\dirR}{\logPhi}
    \\
    \sem{\stepsN}{\distD}{\dirR}{\logPhi_1 \land \logPhi_2}
    &=
    \sem{\stepsN}{\distD}{\dirR}{\logPhi_1}
    \cap
    \sem{\stepsN}{\distD}{\dirR}{\logPhi_2}
    \\
    \sem{\stepsN}{\distD}{\dirR}{\logPr{\rhd \probP}{\logPsi}}
    &=
    \setcomp{\stateQ\in\tsQ}{
    \Pr(\trStart{\stateQ} \cap \sem{\stepsN}{\distD}{\dirR}{\logPsi})
    + \dirR \cdot \distD \rhd \probP }
    \\
    \sem{\stepsN}{\distD}{\dirR}{\logNext \logPhi}
    &=
    \setcomp{\traceT}{t(1) \in \sem{\stepsN}{\distD}{\dirR}{\logPhi}}
    \\
    \sem{\stepsN}{\distD}{\dirR}{\logPhi_1 \logUntil \logPhi_2}
    &=
    \setcomp{\traceT}{
    \exists i\in 0..\stepsN.\
    t(i) \in \sem{\stepsN}{\distD}{\dirR}{\logPhi_2}
    \land
    \forall j\in 0..i-1.\ t(j) \in \sem{\stepsN}{\distD}{\dirR}{\logPhi_1}}
  \end{array}
  \]
\end{defi}

The semantics is mostly standard, except for
$\logPr{\rhd \probP}{\logPsi}$ and $\logPhi_1 \logUntil \logPhi_2$.
The semantics of $\logPr{\rhd \probP}{\logPsi}$ adds
$\dirR\cdot\distD$ to the probability of satisfying $\logPsi$, which
relaxes or strengthens (depending on $\dirR$) the probability bound as
needed.
The semantics of $\logPhi_1 \logUntil \logPhi_2$ uses the parameter
$\stepsN$ to bound the number of steps within which $\logPhi_2$ must
hold.

Our semantics enjoys monotonicity.
The semantics of state and path formulae is increasing \wrt~$\distD$
if $\dirR = +1$, and decreasing otherwise.
The semantics also increases when moving from $\dirR=-1$ to
$\dirR=+1$.

\begin{lem}[Monotonicity]
  \label{lem:pctl:monotonicity}
  Whenever $\distD \leq \distDi$, we have:
  \begin{align*}
    & \sem{\stepsN}{\distD}{+1}{\logPhi} \subseteq
      \sem{\stepsN}{\distDi}{+1}{\logPhi}
    && \sem{\stepsN}{\distDi}{-1}{\logPhi} \subseteq
      \sem{\stepsN}{\distD}{-1}{\logPhi}
    && \sem{\stepsN}{\distD}{-1}{\logPhi} \subseteq
      \sem{\stepsN}{\distD}{+1}{\logPhi}
    \\
    & \sem{\stepsN}{\distD}{+1}{\logPsi} \subseteq
      \sem{\stepsN}{\distDi}{+1}{\logPsi}
    && \sem{\stepsN}{\distDi}{-1}{\logPsi} \subseteq
       \sem{\stepsN}{\distD}{-1}{\logPsi}
    && \sem{\stepsN}{\distD}{-1}{\logPsi} \subseteq
       \sem{\stepsN}{\distD}{+1}{\logPsi}
  \end{align*}
\end{lem}

Note that monotonicity does \emph{not} hold for the parameter $\stepsN$,
\ie even if $\stepsN\leq\stepsNi$, we can \emph{not} conclude
$\sem{\stepsN}{\distD}{+1}{\logPhi} \subseteq
\sem{\stepsNi}{\distD}{+1}{\logPhi}$.
As a counterexample, let
$\tsQ = \setenum{\stateQ_0, \stateQ_1}$, $\lab(\stateQ_0)=\emptyset$,
$\lab(\stateQ_1)=\setenum{\sf a}$,
$\tsPr{\stateQ_0}{\stateQ_1}=\tsPr{\stateQ_1}{\stateQ_1}=1$, and
$\tsPr{\stateQ}{\stateQ'}=0$ elsewhere.
Given $\logPhi=\logPr{\leq 0}{{\sf true}\ \logUntil\ {\sf a}}$, we
have $\stateQ_0 \in \sem{0}{0}{+1}{\logPhi}$ since in $\stepsN=0$
steps it is impossible to reach a state satisfying $\sf a$.
However, we do \emph{not} have $\stateQ_0 \in \sem{1}{0}{+1}{\logPhi}$
since in $\stepsNi=1$ steps we always reach $\stateQ_1$, 
which satisfies $\sf a$.

\begin{figure}[t]
  \centering
  \begin{tikzpicture}[scale=0.9, transform shape, >=latex]
    \node [state] (OK) at (0, 0) {$q_{\sf ok}$};
    \path [->] (OK) edge [loop left] node {\small $1$} (OK);
    \node [state] (Q0) at (2, 0) {$q_{0}$};
    \node [state] (Q1) at (4, 0) {$q_{1}$};
    \draw [->] (Q0) -- (Q1) node[pos=0.5,above] {\small $1-\frac{1}{N}$};
    \node [state] (Q2) at (6, 0) {$q_{2}$};
    \draw [->] (Q1) -- (Q2) node[pos=0.5,above] {\small $1\!-\!\frac{1}{N-1}$};
    \node [state] (Q3) at (8, 0) {$q_{3}$};
    \draw [->] (Q2) -- (Q3) node[pos=0.5,above] {\small $1\!-\!\!\frac{1}{N-2}$};
    \node [state] (ERR) at (2, -2) {$q_{\sf err}$};
    \node [] (Q4) at (9.75, 0) {};
    \draw [->,dotted] (Q3) -- (Q4) node[pos=0.5,above] {\small $1-\frac{1}{N-3}$};
    \node [] (QNp) at (10.25, 0) {};
    \node [state,ellipse] (QN-1) at (11.5, 0) {$q_{N-1}$};
    \draw [->,dotted] (QNp) -- (QN-1) node[pos=0.5,above] {\small $\frac{1}{2}$};
    \path [->] (ERR) edge [loop left] node {\small $1$} (ERR);
    \path [->] (Q0.south) edge node[left] {\small $\frac{1}{N}$} (ERR.north);
    \path [->] (Q1.south) edge [out=-90, in=75] node[above] {\small $\frac{1}{N-1}$} (ERR.north east);
    \path [->] (Q2.south) edge [out=-90, in=45] node[above] {\small $\frac{1}{N-2}$} (ERR.north east);
    \path [->] (Q3.south) edge [out=-90, in=25] node[above] {\small $\frac{1}{N-3}$} (ERR.north east);
    \path [->] (QN-1.south) edge [out=-90, in=15] node[above] {\small $1$} (ERR.north east);
  \end{tikzpicture}
  \\[-40pt]
  \caption{A Markov chain modelling an ideal (left) and a real (right) padlock.}
  \label{fig:padlock}
\end{figure}
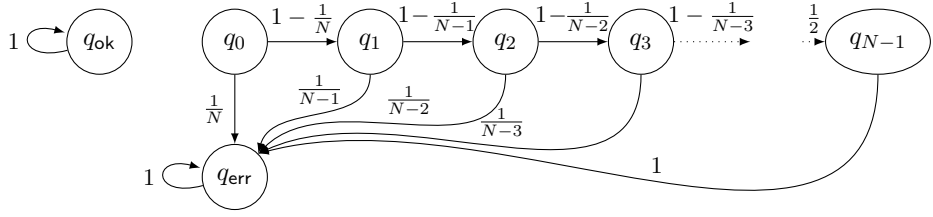

\begin{exa}\label{ex:pctl:padlock}

We compare an ideal combination padlock to a real one from the point
of view of an adversary.
The ideal padlock has a single state $\stateQ_{\sf ok}$, representing
a closed padlock that can not be opened.
Instead, the real padlock is under attack from the adversary who tries
to open the padlock by repeatedly guessing its 5-digit PIN.
At each step the adversary generates a (uniformly) random PIN,
different from all the ones which have been attempted so far, and
tries to open the padlock with it.
The states of the real padlock are ${\stateQ_0,\ldots,\stateQ_{N-1}}$
(with $N=10^5$), where $\stateQ_i$ represents the situation where $i$
unsuccessful attempts have been made, and an additional state
$\stateQ_{\sf err}$ that represents that the padlock was opened.

Since after $i$ attempts the adversary needs to guess the correct PIN
among the $N-i$ remaining combinations, the real padlock in state
$\stateQ_i$ moves to $\stateQ_{\sf err}$ with probability $1/(N-i)$,
and to $\stateQ_{i+1}$ with the complementary probability.

Summing up, we simultaneously model both the ideal and real padlock
as a single DTMC with the following transition probability function
(see~\autoref{fig:padlock}):
\[
  \begin{array}{l@{\qquad}l}
    \Pr(\stateQ_{\sf ok},\stateQ_{\sf ok})=1
    \\
    \Pr(\stateQ_{\sf err},\stateQ_{\sf err})=1
    \\
    \Pr(\stateQ_i,\stateQ_{\sf err}) = 1/(N-i) & 0\leq i<N
    \\
    \Pr(\stateQ_i,\stateQ_{i+1}) = 1-1/(N-i) & 0\leq i<N-1
    \\
    \Pr(\stateQ,\stateQi) = 0 & \text{otherwise}
  \end{array}
\]
We label the states with labels $\labels=\setenum{\sf err}$ by letting
$\lab(\stateQ_{\sf err})=\setenum{\sf err}$ and
$\lab(\stateQ)=\emptyset$ for all $\stateQ\neq\stateQ_{\sf err}$.

The PCTL formula
$\logPhi = \logPr{\leq 0}{{\sf true}\ \logUntil\ {\sf err}}$
models the expected behaviour of an unbreakable padlock, requiring that the
set of traces where the padlock is eventually opened has zero
probability.
Formally, $\logPhi$ is satisfied by state $\stateQ$ when
\begin{align}
  \nonumber
  \stateQ \in \sem{\stepsN}{\distD}{+1}{\logPhi}
  & \iff 
    \stateQ \in \sem{\stepsN}{\distD}{+1}{\lnot \logPr{> 0}{{\sf true}\ \logUntil\ {\sf err}}}
  \\
  \nonumber
  & \iff 
    \stateQ \notin \sem{\stepsN}{\distD}{-1}{\logPr{> 0}{{\sf true}\ \logUntil\ {\sf err}}}
  \\
  \nonumber
  & \iff 
    \lnot ( \Pr(\trStart{\stateQ} \cap \sem{\stepsN}{\distD}{-1}{{\sf true}\ \logUntil\ {\sf err}}) - \distD > 0 )
  \\
  \label{eq:padlock-pr}
  & \iff 
    \Pr(\trStart{\stateQ} \cap \sem{\stepsN}{\distD}{-1}{{\sf true}\ \logUntil\ {\sf err}}) \leq \distD
\end{align}

When $\stateQ=\stateQ_{\sf ok}$ we have that
$\trStart{\stateQ_{\sf ok}} \cap\sem{\stepsN}{\distD}{-1}{{\sf true}\
  \logUntil\ {\sf err}} = \emptyset$, hence the above probability is
zero, which is surely $\leq \distD$.
Consequently, $\logPhi$ is satisfied by the ideal padlock
$\stateQ_{\sf ok}$, for all $\stepsN\geq 0$ and $\distD\geq 0$.

By contrast, $\logPhi$ is not always satisfied by the real padlock
$\stateQ=\stateQ_0$, since we have
$\stateQ_0\in \sem{\stepsN}{\distD}{+1}{\logPhi}$ only for some values
of $\stepsN$ and $\distD$.
To show why, we start by considering some trivial cases.
Choosing $\distD=1$ makes equation~\eqref{eq:padlock-pr}
trivially true for all $\stepsN$.
Furthermore, if we choose $\stepsN=1$,
then $\trStart{\stateQ_0} \cap\sem{\stepsN}{\distD}{-1}{{\sf true}\ \logUntil\ {\sf err}}
= \setenum{\stateQ_0\stateQ_{\sf err}^\omega}$
is a set of traces with probability $1/N$.
Therefore, equation~\eqref{eq:padlock-pr} holds only when
$\distD\geq 1/N$.
More in general, when $\stepsN\geq 1$, we have
\[
\trStart{\stateQ_0} \cap
\sem{\stepsN}{\distD}{-1}{{\sf true}\ \logUntil\ {\sf err}} =
\setenum{\stateQ_0\stateQ_{\sf err}^\omega,\ 
  \stateQ_0\stateQ_1\stateQ_{\sf err}^\omega,\ 
  \stateQ_0\stateQ_1\stateQ_2\stateQ_{\sf err}^\omega,\ 
  \ldots,\ 
  \stateQ_0\ldots\stateQ_{n-1}\stateQ_{\sf err}^\omega
}
\]
The probability of the above set is the probability of guessing the
PIN within $\stepsN$ steps. The complementary event, \ie not guessing
the PIN for $\stepsN$ times, has probability
\[
\dfrac{N-1}{N} \cdot
\dfrac{N-2}{N-1}
\cdots
\dfrac{N-\stepsN}{N-(\stepsN-1)}  =
\dfrac{N-\stepsN}{N}
\]
Consequently, \eqref{eq:padlock-pr} simplifies to
$\stepsN/N \leq \distD$, suggesting the least value of $\distD$
(depending on $\stepsN$) for which $\stateQ_0$ satisfies $\logPhi$.
For instance, when $\stepsN=10^3$, this amounts to claiming that the
real padlock is secure, up to an error of
$\distD = \stepsN/N = 10^{-2}$.
\end{exa}

\section{Up-to-$\stepsN,\distD$ Bisimilarity}

We now define a relation on states
$\stateQ \crysim{\stepsN}{\distD} \stateQi$ that intuitively holds
whenever $\stateQ$ and $\stateQi$ exhibit similar behaviour for
a bounded number of steps.
The parameter $\stepsN$ controls the number of steps, while $\distD$
controls the error allowed in each step.
Note that since we only observe the first $\stepsN$ steps, our notion is
\emph{inductive}, unlike unbounded bisimilarity which is co-inductive,
similarly to~\cite{Castiglioni16qapl}.
Our notion is also inspired by~\cite{Desharnais08qest}.

\begin{defi}[Up-to-$\stepsN,\distD$ Bisimilarity]
  \label{def:param-bisim}
  We define the relation $\stateQ \crysim{\stepsN}{\distD} \stateQi$
  as follows by induction on $\stepsN$:
  \begin{enumerate}
  \item $\stateQ \crysim{0}{\distD} \stateQi$
    always holds
  \item $\stateQ \crysim{\stepsN+1}{\distD} \stateQi$
    holds if and only if, for all $\statesQ \subseteq \tsQ$:
    \begin{enumerate}
    \item\label{def:param-bisim:a}
      $\lab(\stateQ) = \lab(\stateQi)$
    \item\label{def:param-bisim:b}
      $\tsPr{\stateQ}{\statesQ} \leq
      \tsPr{\stateQi}{\cryset{\stepsN}{\distD}{\statesQ}} + \distD$
    \item\label{def:param-bisim:c}
      $\tsPr{\stateQi}{\statesQ} \leq
      \tsPr{\stateQ}{\cryset{\stepsN}{\distD}{\statesQ}} + \distD$
    \end{enumerate}
  \end{enumerate}
  where $\cryset{\stepsN}{\distD}{\statesQ} =
    \setcomp{\stateQi}{\exists \stateQ\in\statesQ.\
      \stateQ \crysim{\stepsN}{\distD} \stateQi}$
    is the image of the set $\statesQ$ according to the bisimilarity
    relation.
\end{defi}

We now establish two basic properties of the bisimilarity.
Our notion is reflexive and symmetric, and enjoys a triangular
property. Furthermore, it is monotonic on both $\stepsN$ and $\distD$.

\begin{lem}
  The relation $\crysim{}{}$ satisfies:
  \[
    \stateQ \crysim{\stepsN}{\distD} \stateQ
    \qquad\quad
    \stateQ \crysim{\stepsN}{\distD} \stateQi
    \implies \stateQi \crysim{\stepsN}{\distD} \stateQ
    \qquad\quad
    \stateQ \crysim{\stepsN}{\distD} \stateQi      
    \land \stateQi \crysim{\stepsN}{\distDi} \stateQii
    \implies \stateQ \crysim{\stepsN}{\distD+\distDi} \stateQii
  \]
\end{lem}
\begin{proof}
  Straightforward induction on $\stepsN$.
\end{proof}

\begin{lem}[Monotonicity]
  \label{lem:sim:monotonicity}
  \begin{align*}
    \stepsNi \leq \stepsN
    & \;\;\implies\;\;
    \crysim{\stepsN}{\distD} \;\;\subseteq\;\; \crysim{\stepsNi}{\distD}
    \\
    \distD \leq \distDi
    & \;\;\implies\;\;
    \crysim{\stepsN}{\distD} \;\;\subseteq\;\;  \crysim{\stepsN}{\distDi}
  \end{align*}
\end{lem}

\begin{exa}
  \label{ex:sim:padlock}
  We use up-to-$\stepsN,\distD$ bisimilarity to compare the behaviour of the
  ideal padlock $\stateQ_{\sf ok}$ and the real one, in any of its
  states, when observed for $\stepsN$ steps.
  When $\stepsN=0$ bisimilarity trivially holds, so below we only
  consider $\stepsN>0$.

  We start from the simplest case: bisimilarity does not hold
  between $\stateQ_{\sf ok}$ and $\stateQ_{\sf err}$.
  Indeed, $\stateQ_{\sf ok}$ and $\stateQ_{\sf err}$ have distinct
  labels
  ($\lab(\stateQ_{\sf ok})=\emptyset\neq\setenum{{\sf
      err}}=\lab(\stateQ_{\sf err})$), hence we do not have
  $\stateQ_{\sf ok} \crysim{\stepsN}{\distD} \stateQ_{\sf err}$, no
  matter what $\stepsN>0$ and $\distD$ are.

  We now compare $\stateQ_{\sf ok}$ with any $\stateQ_i$.
  When $\stepsN=1$, both states have an empty label set,
  i.e.~$\lab(\stateQ_{\sf ok})=\lab(\stateQ_i)=\emptyset$, hence
  they are bisimilar for any error $\distD$.
  We therefore can write $\stateQ_{\sf ok} \crysim{1}{\distD} \stateQ_i$ for any
  $\distD\geq 0$.

  When $\stepsN=2$, we need a larger error $\distD$ to make
  $\stateQ_{\sf ok}$ and $\stateQ_i$ bisimilar.
  Indeed, if we perform a move from $\stateQ_i$, the padlock can be
  broken with probability $1/(N-i)$, in which case we reach
  $\stateQ_{\sf err}$, thus violating bisimilarity.
  Accounting for such probability, we only obtain
  $\stateQ_{\sf ok} \crysim{2}{\distD} \stateQ_i$
  for any $\distD\geq 1/(N-i)$.

  When $\stepsN=3$, we need an even larger error $\distD$ to make
  $\stateQ_{\sf ok}$ and $\stateQ_i$ bisimilar.
  Indeed, while the first PIN guessing attempt has probability
  $1/(N-i)$, in the second move the guessing probability increases to
  $1/(N-i-1)$.
  Choosing $\distD$ equal to the largest probability is enough to
  account for both moves, hence we obtain
  $\stateQ_{\sf ok} \crysim{3}{\distD} \stateQ_i$ for any
  $\distD\geq 1/(N-i-1)$.
  Technically, note that the denominator $N-i-1$ might be zero, since
  when $i=\stepsN-1$ the first move always guesses the PIN, and the
  second guess never actually happens.
  In such case, we instead take $\distD=1$.
  More in detail, we verify item~\eqref{def:param-bisim:b} of
  \autoref{def:param-bisim} for
  $\stateQ_{\sf ok} \crysim{3}{\distD} \stateQ_i$,
  assuming $\distD\geq 1/(N-i-1)$.
  We must prove that:
  \[
  \tsPr{\stateQ_{\sf ok}}{\statesQ} \leq
  \tsPr{\stateQ_i}{\cryset{2}{\distD}{\statesQ}} + \distD
  \]
  When $\stateQ_{\sf ok} \not\in \statesQ$ we have
  $\tsPr{\stateQ_{\sf ok}}{\statesQ} = 0$, hence the inequality holds trivially.
  Otherwise, if $\stateQ_{\sf ok} \in \statesQ$ we first observe that
  $\tsPr{\stateQ_{\sf ok}}{\statesQ} = 1$.
  From the case $\stepsN = 2$, we have
  $\stateQ_{\sf ok} \crysim{2}{\distD} \stateQ_{i+1}$,
  since $\distD \geq 1/(N-(i+1))$.
  Hence, $\stateQ_{i+1} \in \;\cryset{2}{\distD}{\statesQ}$ and so:
  \[
  \tsPr{\stateQ_i}{\cryset{2}{\distD}{\statesQ}} + \distD
  \geq
  \tsPr{\stateQ_i}{\setenum{\stateQ_{i+1}}} + \distD
  =
  1 - \dfrac{1}{N-i} + \distD
  \geq
  1 - \dfrac{1}{N-i} + \dfrac{1}{N-i-1}
  \geq
  1
  \]
  This proves item~\eqref{def:param-bisim:b};
  the proof for item~\eqref{def:param-bisim:c} is similar.
  
  More in general, for an arbitrary $\stepsN\geq 2$, we obtain
  through a similar argument that
  $\stateQ_{\sf ok} \crysim{\stepsN}{\distD} \stateQ_i$
  for any $\distD\geq 1/(N-i-\stepsN+2)$.
  Intuitively, $\distD=1/(N-i-\stepsN+2)$ is the probability of
  guessing the PIN in the last attempt (the $\stepsN$-th), which is
  the attempt having the highest success probability.
  Again, when the denominator $N-i-\stepsN+2$ becomes zero (or
  negative), we instead take $\distD=1$.
  
\end{exa}

Note that the DTMC of the ideal and real padlocks
(Example~\ref{ex:pctl:padlock}) has finitely many states.
Our bisimilarity notion and results, however, can also deal with DTMCs
with a countably infinite set of states, as we show in the next example.

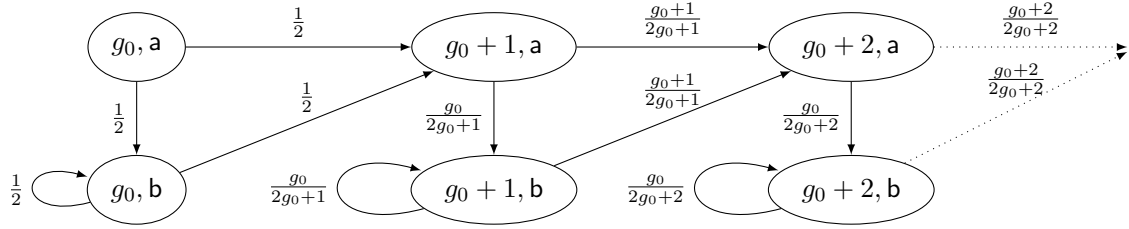
\begin{figure}[t] 
  \centering
  \begin{tikzpicture}[scale=0.95, transform shape, >=latex]
    \node [state,ellipse] (A0) at (4, 0) {$g_{0},{\sf a}$};
    \node [state,ellipse] (A1) at (9, 0) {$g_{0} + 1,{\sf a}$};
    \draw [->] (A0) -- (A1) node[pos=0.5,above] {\small $\frac{1}{2}$};
    \node [state,ellipse] (A2) at (14, 0) {$g_{0} + 2,{\sf a}$};
    \draw [->] (A1) -- (A2) node[pos=0.5,above] {\small $\frac{g_0 + 1}{2 g_0 + 1}$};
    \node (A3) at (18, 0) {};
    \draw [->,dotted] (A2) -- (A3) node[pos=0.5,above] {\small $\frac{g_0+2}{2 g_0 + 2}$};
    \node [state,ellipse] (B0) at (4, -2) {$g_{0},{\sf b}$};
    \path [->] (B0) edge [loop left] node {\small $\frac{1}{2}$} (B0);
    \path [->] (A0.south) edge node[left] {\small $\frac{1}{2}$} (B0.north);
    \node [state,ellipse] (B1) at (9, -2) {$g_{0}+1,{\sf b}$};
    \path [->] (B1) edge [loop left] node {\small $\frac{g_0}{2 g_0 + 1}$} (B1);
    \draw [->] (B0) -- (A1) node[pos=0.5,above] {\small $\frac{1}{2}$};
    \node [state,ellipse] (B2) at (14, -2) {$g_{0}+2,{\sf b}$};
    \path [->] (B2) edge [loop left] node {\small $\frac{g_0}{2 g_0 + 2}$} (B2);
    \draw [->] (B1) -- (A2) node[pos=0.5,above] {\small $\frac{g_0 + 1}{2 g_0 + 1}$};
    \draw [->] (A1) -- (B1) node[pos=0.5,left] {\small $\frac{g_0}{2 g_0 + 1}$};
    \node (B3) at (18, -2) {};
    \draw [->,dotted] (B2) -- (A3) node[pos=0.5,above] {\small $\frac{g_0+2}{2 g_0 + 2}$};
    \draw [->] (A2) -- (B2) node[pos=0.5,left] {\small $\frac{g_0}{2 g_0 + 2}$};
  \end{tikzpicture}
  \caption{A Markov chain modelling an unfair random generator of bit streams.} 
  \label{fig:pingpong}
\end{figure}

\begin{exa}
  \label{ex:sim:pingpong}
  We consider an ideal system which randomly generates bit streams
  in a fair way.
  We model such a system as having two states
  $\setenum{\stateQ_a,\stateQ_b}$, with transition probabilities
  $\Pr(x,y)=1/2$ for any $x,y \in \setenum{\stateQ_a,\stateQ_b}$,
  as in~\autoref{fig:fair-coin}.
  We label state $\stateQ_a$ with label $\sf a$ denoting bit $0$, and
  state $\stateQ_b$ with label $\sf b$ denoting bit $1$.

  We compare this ideal system with a real system which generates
  bit streams in an unfair way.
  At each step, the real system draws a ball from an urn, initially
  having $g_0$ $\sf a$-labelled balls and $g_0$ $\sf b$-labelled
  balls.
  After each drawing, the ball is placed back in the urn.
  However, every time an $\sf a$-labelled ball is drawn, an additional
  $\sf a$-labelled ball is put in the urn, making the next drawings
  more biased towards $\sf a$.

  We model the real system using the infinite\footnote{%
    Modelling this behaviour inherently requires an \emph{infinite}
    set of states, since each number of $\sf a$-labelled balls in the urn
    leads to a unique transition probability function.} set of states
  $\mathbb N \times \setenum{{\sf a},{\sf b}}$, whose first component
  counts the number of $\sf a$-labelled balls in the urn, and the
  second component is the label of the last-drawn ball.
  The transition probabilities are as follows, where $g_0\in\mathbb N^+$
  (see~\autoref{fig:pingpong}):
  \[
    \begin{array}{ll@{\qquad}l}
      \Pr((g,x),(g+1,{\sf a})) &= g / (g+g_0)
      \\
      \Pr((g,x),(g,{\sf b})) &= g_0 / (g+g_0)
      \\
      \Pr((g,x),(g',x')) &= 0 & \text{otherwise}
    \end{array}
  \]
  We label each such state with its second component.

  We now compare the ideal system to the real one.
  Intuitively, the ideal system, when started from state $\stateQ_a$,
  produces a sequence of states whose labels are uniform independent
  random values in $\setenum{{\sf a},{\sf b}}$.
  Instead, the real system slowly becomes more and more biased towards
  label $\sf a$.
  More precisely, when started from state $(g_0, {\sf a})$, in the
  first drawing the next label is uniformly distributed between
  ${\sf a}$ and ${\sf b}$, as in the ideal system.
  When the sampled state has label $\sf a$, this causes the component
  $g$ to be incremented, increasing the probability $g/(g+g_0)$ of
  sampling another $\sf a$ in the next steps.
  Indeed, the value $g$ is always equal to $g_0$ plus the number of
  sampled $\sf a$-labelled states so far.
  
  Therefore, unlike the ideal system, on the long run the real system
  will visit $\sf a$-labelled states with very high probability, since
  the $g$ component slowly but steadily increases.
  While this fact makes the two systems \emph{not} bisimilar according
  to the standard probabilistic bisimilarity~\cite{Larsen89popl}, if
  we restrict the number of steps to $\stepsN \ll g_0$ and tolerate a
  small error $\distD$, we can obtain
  $\stateQ_a \crysim{\stepsN}{\distD} (g_0,{\sf a})$.

  For instance, if we let $g_0=1000$, $\stepsN=100$ and $\delta=0.05$
  we have $\stateQ_a \crysim{\stepsN}{\distD} (g_0,{\sf a})$.
  This is because, in $\stepsN$ steps, the first component $g$ of a
  real system $(g,x)$ will at most reach $1100$, making the
  probability of the next step to be $(g+1,{\sf a})$ to be at most
  $1100/2100\simeq 0.523$.
  This differs from the ideal probability $0.5$ by less than $\distD$,
  hence bisimilarity holds.
\end{exa}

\section{Soundness}
\label{sec:result}

Our soundness theorem shows that, if we consider any state $\stateQ$
satisfying $\logPhi$ (with steps $\stepsN$ and error $\distDi$), and
any state $\stateQi$ which is bisimilar to $\stateQ$ (with enough
steps and error $\distD$), then $\stateQi$ must satisfy $\logPhi$,
with the same number $\stepsN$ of steps, at the cost of suitably
increasing the error.
For a fixed $\logPhi$, the ``large enough'' number of steps and the
increase in the error depend linearly on $\stepsN$.

\begin{thm}[Soundness]
  \label{th:soundness}
  Let $\boundK_X = \nestMax{\logNext}{\logPhi}$ 
  be the maximum $\logNext$-nesting of a formula $\logPhi$, 
  and let $\boundK_U = \nestMax{\logUntil}{\logPhi}$ 
  be the maximum $\logUntil$-nesting of $\logPhi$.
  Then, for all $\stepsN,\distD,\distDi$ we have:
  \[
    \begin{array}{c}
      \cryset{\stepsNb}{\distD}{
      \sem{\stepsN}{\distDi}{+1}{\logPhi}}
      \subseteq
      \sem{\stepsN}{\stepsNb\cdot\distD+\distDi}{+1}{\logPhi}
      \tag*{ where $\stepsNb = \stepsN\cdot\boundK_U + \boundK_X + 1$}      
    \end{array}
  \]
\end{thm}

\begin{exa}
  \label{ex:results:padlock}
  We apply~\autoref{th:soundness} to our padlock system
  in the running example.
  We take the same formula
  $\logPhi = \logPr{\leq 0}{{\sf true}\ \logUntil\ {\sf err}}$
  of~\autoref{ex:pctl:padlock} and choose $\stepsN=10^3$ and $\distDi=0$.
  Since $\logPhi$ has only one until operator and no next operators,
  the value $\stepsNb$ in the theorem statement is
  $\stepsNb = 10^3\cdot 1+0+1 = 1001$.
  Therefore, from~\autoref{th:soundness} we obtain, for all $\distD$:
  \[
    \begin{array}{ll}
      & \cryset{1001}{\distD}{
        \sem{1000}{0}{+1}{\logPhi}}
        \subseteq
        \sem{1000}{1001\cdot \distD}{+1}{\logPhi}
    \end{array}
  \]
  
  In~\autoref{ex:pctl:padlock} we discussed how the ideal padlock
  $\stateQ_{\sf ok}$ satisfies the formula $\logPhi$ for any
  number of steps and any error value.
  In particular, choosing 1000 steps and zero error, we get
  $\stateQ_{\sf ok}\in \sem{1000}{0}{+1}{\logPhi}$.
  
  Moreover, in~\autoref{ex:sim:padlock} we observed that states
  $\stateQ_{\sf ok}$ and $\stateQ_0$ are bisimilar with
  $\stepsNb=1001$ and $\distD=1/(N-0-\stepsNb+2) = 1/99001$,
  \ie~$\stateQ_{\sf ok} \crysim{\stepsNb}{\distD} \stateQ_0$.

  In such case, the theorem ensures that
  $\stateQ_0\in\sem{1000}{1001/99001}{+1}{\logPhi}$, hence the
  real padlock can be considered unbreakable if we limit our
  attention to the first $\stepsN=1000$ steps, up to an error
  of $1001/99001 \approx 0.010111$.
  Finally, we note that such error is remarkably close to the least
  value that would still make $\stateQ_0$ satisfy $\logPhi$, which we
  computed in~\autoref{ex:pctl:padlock} as
  $\stepsN/N = 10^3/10^5 = 0.01$.
\end{exa}

In the rest of this section, we describe the general structure of the
proof in a top-down fashion, leaving the detailed proof
for~\autoref{sec:proofs}.

We prove the soundness theorem by induction on the state formula
$\logPhi$, hence we also need to deal with path formulae $\logPsi$.
Note that the statement of the theorem considers the image of the
semantics of the state formula $\logPhi$ w.r.t.~bisimilarity (i.e.,
$\cryset{\stepsNb}{\distD}{\sem{\stepsN}{\distDi}{+1}{\logPhi}}$).
Analogously, to deal with path formulae we also need an analogous
notion on sets of traces.
To this purpose, we consider the set of traces in the definition of
the semantics:
$\tracesT = \trStart{\stateP} \cap
\sem{\stepsN}{\distD}{\dirR}{\logPsi}$.
Then, given a state $\stateQ$ bisimilar to $\stateP$, we define the
set of \emph{pointwise bisimilar traces} starting from $\stateQ$,
which we denote with $\TR{\stepsN}{\distD,\sQ}{\tracesT}$.
Technically, since $\logPsi$ can only observe a finite portion of a
trace, it is enough to define $\TR{\stepsN}{\distD,\sQ}{\finTracesT}$
on sets of \emph{trace fragments} $\finTracesT$.

\begin{defi}
  Write $\fram{\stateQ_0}{\stepsN}$ for the set of all trace fragments
  of length $n$ starting from $\stateQ_0$.
  Assuming $\sP \CSim{\stepsN}{\distD} \sQ$,
  we define $\TR{\stepsN}{\distD,\sQ}{}: \powset(\fram{\sP}{\stepsN})
  \rightarrow \powset(\fram{\sQ}{\stepsN})$
  as follows:
  \[
  \TR{\stepsN}{\distD,\sQ}{\finTracesT} =
  \setcomp{\finTraceU \in \fram{\sQ}{\stepsN}}{
    \exists \finTraceT \in \finTracesT.\, \forall 0 \leq i < \stepsN.\ 
    \finTraceT(i) \CSim{\stepsN-i}{\distD} \finTraceU(i)
  }
  \]
\end{defi}

\noindent
In particular, notice that
$\fram{\stateQ}{1} = \setenum{\stateQ}$ (the trace fragment of length 1),
and so:
\[
\TR{1}{\distD,\sQ}{\emptyset} = \emptyset
\qquad
\TR{1}{\distD,\sQ}{\setenum{\stateQ}} = \setenum{\stateQ}
\]

The key inequality we exploit in the proof (\autoref{lem:traces})
compares the probability of a set of trace fragments $\finTracesT$
starting from $\sP$ to the one of the related set of trace fragments
$\TR{m}{\distD,\sQ}{\finTracesT}$ starting from a $\sQ$ bisimilar to
$\sP$.
We remark that the component $\stepsNb \distD$ in the error that
appears in~\autoref{th:soundness} results from the component $m \distD$
appearing in the following lemma.

\begin{lem}\label{lem:traces}
  If $\sP \CSim{\stepsN}{\distD} \sQ$ and $\finTracesT$ is a set of
  trace fragments of length $m$, with $m \leq \stepsN$, starting
  from $\sP$, then:
\[
  \prob{\finTracesT}{} \leq \prob{\TR{m}{\distD,\sQ}{\finTracesT}}{} +
  m \distD
\]
\end{lem}

\autoref{lem:traces} allows $\finTracesT$ to be an infinite set (because
the set of states $\tsQ$ can be infinite).
We reduce this case to that in which $\finTracesT$ is finite.
We first recall a basic calculus property: any inequality $a \leq b$
can be proved by establishing instead $a \leq b + \epsilon$ for all
$\epsilon > 0$.
Then, since the probability distribution of trace fragments of length
$m$ is discrete, for any $\epsilon>0$ we can always take a finite
subset of the infinite set $\finTracesT$ whose probability differs
from that of $\finTracesT$ less than $\epsilon$.
It is then enough to consider the case in which $\finTracesT$ is
finite, as done in the following lemma.

\begin{lem}\label{lem:finite-traces}
  If $\sP \CSim{\stepsN}{\distD} \sQ$ and $\finTracesT$ is a finite
  set of trace fragments of length $\stepsN > 0$ starting from $\sP$,
  then:
\[
  \prob{\finTracesT}{} \leq
  \prob{\TR{\stepsN}{\distD,\sQ}{\finTracesT}}{} + \stepsN \distD
\]
\end{lem}

We prove~\autoref{lem:finite-traces} by induction on $\stepsN$.  In the
inductive step, we partition the traces according to their first move,
i.e., on their next state after $\sP$ (for the trace fragments in $T$)
or $\sQ$ (for the bisimilar counterparts).
A main challenge here is caused by the probabilities of such moves
being weakly connected. Indeed, when $\sP$ moves to $\sPi$, we might
have several states $\sQi$, bisimilar to $\sPi$, such that $\sQ$ moves
to $\sQi$. Worse, when $\sP$ moves to another state $\sPii$, we might
find that some of the states $\sQi$ we met before are also bisimilar
to $\sPii$.
Such overlaps make it hard to connect the probability of $\sP$ moves
to that of $\sQ$ moves.

To overcome these issues, we exploit the technical lemma below.  Let
set $A$ represent the $\sP$ moves, and set $B$ represent the $\sQ$
moves.
Then, associate to each set element $a\in A,b\in B$ a value
($f_A(a), f_B(b)$ in the lemma) representing the move probability.
The lemma ensures that each $f_A(a)$ can be expressed as a weighted
sum of $f_B(b)$ for the elements $b$ bisimilar to $a$.  Here, the
weights $h(a,b)$ make it possible to relate a $\sP$ move to a
``weighted set'' of $\sQ$ moves.
Furthermore, the lemma ensures that no $b\in B$ has been cumulatively used
for more than a unit weight ($\sum_{a \in A} h(a,b) \leq 1$).

\begin{lem}\label{lem:matching}
Let $A$ be a finite set and $B$ be a countable set, equipped with functions $f_A: A \rightarrow \NNR$ and $f_B: B \rightarrow \NNR$.
Let $g:A \rightarrow 2^B$ be such that $\sum_{b \in g(a)} f_B(b)$
converges for all $a \in A$. 
If, for all $A' \subseteq A:$
\begin{equation}\label{eq:matching-assumption}
\sum_{a \in A'} f_A(a) \leq \sum_{b \in \bigcup_{a \in A'} g(a)} f_B(b)
\end{equation}
then there exists $h:A \times B \rightarrow \intervalCC 0 1$ such that: 
\begin{align}\label{eq:matching-thesis:1}
  &\forall b \in B: \sum_{a \in A} h(a,b) \leq 1
  \\
  \label{eq:matching-thesis:2}
  &\forall A' \subseteq A: 
  \sum_{a \in A'} f_A(a) = \sum_{a \in A'} \sum_{b \in g(a)} h(a,b) f_B(b)
\end{align}
\end{lem}

\begin{figure}[h]
  \hspace{-5mm}%
  \includegraphics[scale=0.7]{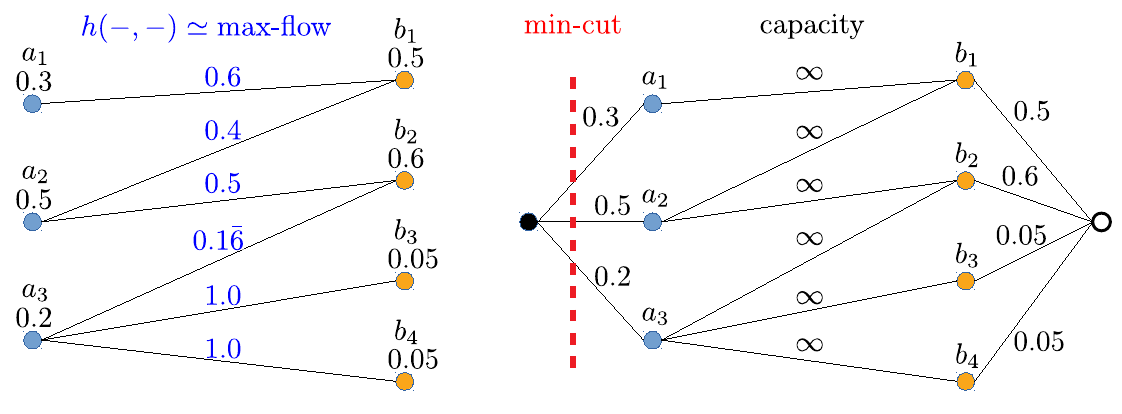}
  \caption{Graphical representation of~\autoref{lem:matching} (left) and
    its proof (right).}
  \label{fig:matching}
\end{figure}

We visualize~\autoref{lem:matching} in~\autoref{fig:matching} through an
example.  The leftmost graph shows a finite set
$A=\setenum{a_1,a_2,a_3}$ where each $a_i$ is equipped with its
associated value $f_A(a_i)$ and, similarly, a finite set
$B=\setenum{b_1,\ldots,b_4}$ where each $b_i$ has its own value
$f_B(b_i)$. The function $g$ is rendered as the edges of the graph,
connecting each $a_i$ with all $b_j \in g(a_i)$.

The graph satisfies the hypotheses, as one can easily verify. For
instance, when $A' = \setenum{a_1,a_2}$ inequality
\eqref{eq:matching-assumption} simplifies to $0.3+0.5 \leq
0.5+0.6$. The thesis ensures the existence of a weight function
$h(-,-)$ whose values are shown in the graph on the left over each
edge.

These values indeed satisfy \eqref{eq:matching-thesis:1}: for
instance, if we pick $b=b_2$ the inequality reduces to
$0.5+0.1\bar 6 \leq 1$. Furthermore, \eqref{eq:matching-thesis:2} is also
satisfied: for instance, taking $A'=\setenum{a_2}$ the equation
reduces to $0.5 = 0.4\cdot 0.5+0.5\cdot 0.6$, while taking
$A'=\setenum{a_3}$ the equation reduces to
$0.2 = 0.1\bar 6\cdot 0.6+1.0\cdot 0.05+1.0\cdot 0.05$.

The rightmost graph in~\autoref{fig:matching} instead sketches how our
proof devises the desired weight function $h$, by constructing a
network flow problem, and exploiting the well-known min-cut/max-flow
theorem~\cite{MinCut}, following the approach of~\cite{Baier98thesis}.
We start by adding a source node to the right (white bullet in the
figure), connected to nodes in $B$, and a sink node to the left,
connected to nodes in $A$.
We write the capacity over each edge: we use $f_B(b_i)$ for the edges
connected to the source, $f_A(a_i)$ for the edges connected to the
sink, and $+\infty$ for the other edges in the middle.

Then, we argue that the leftmost cut $C$ shown in the figure is a
min-cut.  Intuitively, if we take another cut $C'$ not including some
edge in $C$, then $C'$ has to include other edges making $C'$ not any
better than $C$.
Indeed, $C'$ can surely not include any edge in the middle, since they
have $+\infty$ capacity.
Therefore, if $C'$ does not include an edge from some $a_i$ to the
sink, it has to include all the edges from the source to each
$b_j \in g(a_i)$.
In this case, hypothesis \eqref{eq:matching-assumption} ensures that
doing so does not lead to a better cut.
Hence, $C$ is indeed a min-cut.

From the max-flow corresponding to the min-cut, we derive the values
for $h(-,-)$.
Thesis \eqref{eq:matching-thesis:1} follows from the flow conservation
law on each $b_i$, and the fact that the incoming flow of each $b_j$
from the source is bounded by the capacity of the related
edge.
Thesis \eqref{eq:matching-thesis:2} follows from the flow
conservation law on each $a_i$, and the fact that the outgoing flow of
each $a_i$ to the sink is exactly the capacity of the related edge,
since the edge is on a min-cut.

\section{Asymptotic equivalence}
\label{sec:asymptotic}

In this section we transport the notion of bisimilarity and the
semantics of PCTL to \emph{families} of states, thus reasoning on
their asymptotic behaviours.
More precisely, given a state-labelled DTMC $\tsQ$, we define a family
of states to be an infinite sequence $\famF: \numN\to\tsQ$.
Intuitively, $\famF(\eta)$ can describe the behaviour of a
probabilistic system depending on a security parameter $\eta \in \numN$.

When using bisimilarity (\autoref{def:param-bisim}) to relate
two given states $\statesQ_1$ and $\statesQ_2$, we have to provide a
number of steps $\stepsN$ and a probability error $\distD$.
By contrast, when relating two families $\famF_1$ and $\famF_2$ we
want to focus on their asymptotic behaviour, and obtain an equivalence
that does not depend on specific values of $\stepsN$ and
$\distD$.
To do so, we start by recalling the standard definition of
\emph{negligible function}:
\begin{defi}[Negligible Function]
  A function $f : \numN\to\numR$ is said to be negligible
  whenever
  \[
    \forall c\in\numN.\ 
    \exists \bar \eta.\ 
    \forall \eta\geq\bar \eta.\ 
    |f(\eta)| \leq \eta^{-c}
  \]
\end{defi}

We say that $\famF_1$ and $\famF_2$ are asymptotically equivalent
($\famF_1 \simFam \famF_2$) when the families are asymptotically
pointwise bisimilar with a negligible error $\distD(\eta)$ whenever
$\stepsN(\eta)$ is a polynomial.

\begin{defi}[Asymptotic Equivalence]\label{def:crysim}
  Given $\famF_1,\famF_2 : \numN\to\tsQ$, we write
  $\famF_1 \simFam \famF_2$ if and only if
  for each polynomial $\stepsN(-)$
  there exists a negligible function $\distD(-)$
  and $\bar \eta \in \numN$
  such that for all $\eta\geq \bar \eta$
  we have
  $\famF_1(\eta) \crysim{\stepsN(\eta)}{\distD(\eta)} \famF_2(\eta)$
\end{defi}

\begin{lem}
  $\simFam$ is an equivalence relation.
\end{lem}
\begin{proof}
  Reflexivity and symmetry are trivial.  For transitivity, given
  a polynomial
  $\stepsN(-)$, let $\distD_1(-),\distD_2(-)$ be the negligible
  functions resulting from the hypotheses $\famF_1 \simFam \famF_2$
  and $\famF_2 \simFam \famF_3$, respectively.
  Asymptotically, we obtain
  \[
    \famF_1(\eta) \crysim{\stepsN(\eta)}{\distD_1(\eta)} \famF_2(\eta)
    \qquad\land\qquad
    \famF_2(\eta) \crysim{\stepsN(\eta)}{\distD_2(\eta)} \famF_3(\eta)
  \]
  By the transitivity of $\crysim{}{}$, we get
  \[
    \famF_1(\eta)
    \crysim{\stepsN(\eta)}{\distD_1(\eta)+\distD_2(\eta)}
    \famF_3(\eta)
  \]
  Hence we obtain the thesis since the sum of negligible functions
  $\distD_1(\eta)+\distD_2(\eta)$ is negligible.
\end{proof}

We now provide an asymptotic semantics for PCTL, by defining its
satisfaction relation $\famF \models \logPhi$.
As done above, this notion does not depend on specific values for
$\stepsN$ and $\distD$ (unlike the semantics in
\autoref{def:pctl:sem}), but instead considers the asymptotic
behaviour of the family.

\begin{defi}[Asymptotic Satisfaction Relation]%
  We write $\famF \models \logPhi$ when there exists a polynomial
  $\bar\stepsN(-)$ such that for each polynomial
  $\stepsN(-) \geq \bar\stepsN(-)$ there exists a negligible function
  $\distD(-)$ and $\bar\eta \in \numN$ such that for all
  $\eta\geq \bar\eta$ we have
  $\famF(\eta) \in \sem{\stepsN(\eta)}{\distD(\eta)}{+1}{\logPhi}$
\end{defi}

In the above definition, we only consider polynomials greater than a
threshold $\bar\stepsN(-)$.
This is because a family $\famF$ representing, say, a protocol could
require a given (polynomial) number of steps to complete its execution.
It is reasonable, for instance, that $\famF(\eta)$ needs to exchange
$\eta^2$ messages over a network to perform its task.
In such cases, we still want to make $\famF$ satisfy a formula
$\logPhi$ stating that the task is eventually completed with high
probability.
If we quantified over all polynomials $\stepsN(-)$, we would also
allow choosing small polynomials like $\stepsN(\eta)=\eta$ or even
$\stepsN(\eta)=1$, which would not provide $\famF$ enough time to complete.
Using a (polynomial) threshold $\bar\stepsN(-)$, instead, we always
provide enough time.

We now establish the main result of this section, asymptotic
soundness, stating that equivalent families of states asymptotically
satisfy the same PCTL formulae.
The proof relies on our previous soundness~\autoref{th:soundness}.

\begin{thm}[Asymptotic Soundness]
  \label{th:asymptotic}
  Let $\famF_1,\famF_2$ be families of states such that
  $\famF_1 \simFam \famF_2$.  For every PCTL formula $\logPhi$:
  \[
    \famF_1 \models \logPhi \iff
    \famF_2 \models \logPhi
  \]
\end{thm}
\begin{proof}

  Assuming $\famF_1 \models \logPhi$ and
  $\famF_1 \simFam \famF_2$, we prove
  $\famF_2 \models \logPhi$.
  Let $\boundK_X = \nestMax{\logNext}{\logPhi}$ 
  be the maximum $\logNext$-nesting of $\logPhi$, 
  and let $\boundK_U = \nestMax{\logUntil}{\logPhi}$ 
  be the maximum $\logUntil$-nesting of $\logPhi$.

  Let $\bar\stepsN_1(-)$ as in the definition of the hypothesis
  $\famF_1 \models \logPhi$.
  To prove the thesis $\famF_2 \models \logPhi$, we choose
  $\bar\stepsN_2(-) = \bar\stepsN_1(-)$, and we consider an arbitrary
  $\stepsN_2(-)\geq\bar\stepsN_2(-)=\bar\stepsN_1(-)$.  We can then
  choose $\stepsN_1(-) = \stepsN_2(-)$ in the same hypothesis, and
  obtain a negligible $\distD_1(-)$ and $\bar\eta_1$, where for any
  $\eta \geq \bar\eta_1$ we have
  \begin{equation}
    \label{eq:fam1-hp}
    \famF_1(\eta) \in \sem{\stepsN_2(\eta)}{\distD_1(\eta)}{+1}{\logPhi}
  \end{equation}
  
  We now exploit the other hypothesis $\famF_1 \simFam \famF_2$,
  choosing the polynomial
  \begin{equation}
    \label{eq:n-eta}
    \stepsN(\eta) = \stepsN_2(\eta) \cdot \boundK_U + \boundK_X + 1
  \end{equation}
  and obtaining a negligible $\distD(-)$ and $\bar\eta$ where
  for any $\eta\geq\bar\eta$ we have
  \begin{equation}
    \label{eq:sim-hp}
    \famF_1(\eta) \crysim{\stepsN(\eta)}{\distD(\eta)} \famF_2(\eta)
  \end{equation}
  
  To prove the thesis, we finally choose the negligible function
  $\distD_2(\eta) = \stepsN(\eta)\cdot\distD(\eta)+\distD_1(\eta)$ and
  $\bar\eta_2 = \max(\bar\eta_1, \bar\eta)$.
  By~\autoref{th:soundness} we have that for any
  $\eta\geq\bar\eta_2$:
  \[
    \cryset{\stepsN(\eta)}{\distD(\eta)}{
      \sem{\stepsN_2(\eta)}{\distD_1(\eta)}{+1}{\logPhi}}
    \subseteq
    \sem{\stepsN_2(\eta)}{\stepsN(\eta)\cdot\distD(\eta)+\distD_1(\eta)}{+1}{\logPhi}
    \mbox{ where $\stepsN(\eta)$ is as in \eqref{eq:n-eta}.}
  \]
  Applying this to \eqref{eq:fam1-hp} and \eqref{eq:sim-hp} we then
  have that, for any $\eta\geq\bar\eta_2$:
  \[
    \famF_2(\eta) \in \sem{\stepsN_2(\eta)}{\stepsN(\eta)\cdot\distD(\eta)+\distD_1(\eta)}{+1}{\logPhi}
  \]
  which is our thesis
  \[
  \famF_2(\eta) \in \sem{\stepsN_2(\eta)}{\distD_2(\eta)}{+1}{\logPhi}
  \qedhere
  \]
\end{proof}

\begin{exa}\label{ex:asymptotic:padlock}
  We now return to the padlock examples \ref{ex:pctl:padlock} and
  \ref{ex:sim:padlock}.
  We again consider an ideal padlock modelled by a state
  $\stateQ_{\sf ok}$, but also a sequence of padlocks
  having an increasing number of digits $\eta$, hence an increasing
  number $N=10^\eta$ of combinations.
  We assume that state $\stateQ_{i,10^\eta}$ models the state of a
  padlock having $\eta$ digits where the adversary has already made
  $i$ brute force attempts, following the same strategy as in the
  previous examples.
  The transition probabilities are also similarly defined.

  In this scenario, we can define two state families.
  Family $\famF_1(\eta) = \stateQ_{\sf ok}$ represents a (constant)
  sequence of ideal padlocks, while family
  $\famF_2(\eta) = \stateQ_{0,10^\eta}$ represents a sequence of
  realistic padlocks with no previous brute force attempt ($i=0$),
  in increasing order of robustness.
  Indeed, as $\eta$ increases, the padlock becomes harder
  to break by brute force since the number of combinations $N=10^\eta$ grows.

  In~\autoref{ex:sim:padlock}, we have seen that
  \[
    \famF_1(\eta) \crysim{\stepsN(\eta)}{\distD(\eta)} \famF_2(\eta)
    \qquad
    \mbox{ where }
    \delta(\eta) =
    \dfrac{1}{N-0-\stepsN(\eta)+2} =
    \dfrac{1}{10^\eta-\stepsN(\eta)+2}
  \]
  and we can observe that the above $\distD(\eta)$ is indeed
  negligible when $\stepsN(\eta)$ is a polynomial.
  This means that $\famF_1 \simFam \famF_2$ holds, hence we can apply
  \autoref{th:asymptotic} and conclude that the families $\famF_1$
  and $\famF_2$ asymptotically satisfy the same PCTL formulae.
  This is intuitive since, when the adversary can only attempt a
  polynomial number of brute force attacks, and when the number of
  combinations increases exponentially, the robustness of the
  realistic padlocks effectively approaches that of the ideal one.
  
\end{exa}

We now discuss how~\autoref{th:asymptotic} could be applied to
a broad class of systems.
Consider the execution of an ideal cryptographic protocol, modelled
as a DTMC starting from the initial state $\stateQ_i$.
This model could represent, for instance, the semantics of a formal,
symbolic system such as those that can be expressed using process
algebras.
In this scenario, the underlying cryptographic primitives can be
\emph{perfect}, in the sense that ciphertexts reveal no information
to whom does not know the decryption key, signatures can never be
forged, hash preimages can never be found, and so on, despite the
amount of computational resources available to the adversary.

Given such a model, it is then possible to refine it making the
cryptographic primitives more realistic, allowing an adversary to
attempt attacks such as decryptions and signature forgeries, which
however succeed only with negligible probability \wrt a security
parameter $\eta$.
This more realistic system can be modelled using a distinct DTMC
state $\stateQ^\eta_r$ whose behaviour is similar to that of
$\stateQ_i$: the state transitions are essentially the same, except
for the cases in which the adversary is successful in attacking the
cryptographic primitives.
Therefore, the transition probabilities are almost the same,
differing only by a negligible quantity.

Therefore, we can let $\famF_1(\eta)=\stateQ_i$ and
$\famF_2(\eta)=\stateQ^\eta_r$, and observe that they are indeed
asymptotically equivalent.
Note that this holds in general by construction, no matter what is
the behaviour of the ideal system $\stateQ_i$ we started from.
  
Finally, by~\autoref{th:asymptotic} we can claim that both
families $\famF_1,\famF_2$ asymptotically satisfy the same PCTL
formulas.
This makes it possible, in general, to prove properties on the
simpler $\stateQ_i$ system, possibly even using some automatic
verification tools, and transfer these results in the more realistic
setting $\stateQ^\eta_r$.

A special case of this fact was originally studied
in~\cite{ZuninoD05}, which however only considered reachability
properties.
By comparison, \autoref{th:asymptotic} is much more general,
allowing one to transfer any property that can be expressed using a
PCTL formula.

The construction above allows one to refine an ideal system
$\stateQ_i$ into a more realistic one $\stateQ^\eta_r$ by taking
certain adversaries into account.
However, if our goal were to study the security of the system against
\emph{all} reasonable adversaries, then the above approach would not
be applicable.
Indeed, it is easy to find an ideal system and a corresponding
realistic refinement, comprising a reasonable adversary,
where the asymptotic equivalence does not hold.
For instance, consider an ideal protocol where Alice and Bob exchange
ten messages, after which Alice randomly chooses and exchanges a
single bit.
To assess the security of a realistic implementation, we might want to
consider the case where Alice is an adversary.
In such case, a malicious Alice could exchange the first two messages,
then flip a coin $b \leftarrow \{0,1\}$ in secret, exchange the other
eight messages, and finally send the value $b$.
The behaviour of such realistic system differs from the ideal one,
since the ideal one has a probabilistic choice point only at the end,
while the realistic system anticipates it after the first two moves.
It is easy to check (and well known) that moving choices to an earlier
point makes standard bisimilarity fail, and this is the case also for
asymptotic equivalence.
The failure of asymptotic equivalence prevents us from applying the
asymptotic soundness theorem.
In particular, assume that we have proved that the ideal system enjoys
certain specifications expressed as PCTL formulae.
We can not exploit the theorem to show that also the realistic system
with the adversary enjoys the same specifications.

\section{Conclusions}

In this paper we studied how the (relaxed) semantics of PCTL formulae
interacts with (approximate) probabilistic bisimulation. In the
regular, non relaxed case, it is well-known that when a state
$\stateQ$ satisfies a PCTL formula $\logPhi$, then all the states that
are probabilistic-bisimilar to $\stateQ$ also satisfy $\logPhi$
(\cite{Desharnais10iandc}).
\autoref{th:soundness} extends this to the relaxed semantics, establishing that when a state $\stateQ$ satisfies a PCTL formula $\logPhi$ up-to $\stepsN$ steps and error $\distD$, then all the states that are approximately probabilistic bisimilar to $\stateQ$ with error $\distDi$ (and enough steps) also satisfy $\logPhi$ up-to $\stepsN$ steps and suitably increased error. 
We provide a way to compute the new error in terms of $\stepsN, \distD, \distDi$.
\autoref{th:asymptotic} extends such soundness result to the
asymptotic behaviour where the error becomes negligible when the
number of steps is polynomially bounded.

Our results are a first step towards a novel approach to the security
analysis of cryptographic protocols using probabilistic bisimulations.
When one is able to prove that a real-world specification of a
cryptographic protocol is asymptotically equivalent to an ideal one,
then one can invoke~\autoref{th:asymptotic} and claim that the two
models satisfy the same PCTL formulae, essentially reducing the
security proof of the cryptographic protocol to verifying the ideal
model.  A relevant line for future work is to study the applicability
of our theory in this setting.
As discussed in~\autoref{sec:asymptotic}, our approach is not
applicable to all protocols and all adversaries.
A relevant line of research could be the study of larger asymptotic
equivalences, which allow to transfer properties from ideal to
realistic systems.
This could be achieved, e.g., by considering weaker logics than PCTL,
or moving to linear temporal logics. 

Another possible line of research would be investigating proof
techniques for establishing approximate bisimilarity and
refinement~\cite{Jonsson91lics}, as well as devising algorithms for
approximate bisimilarity, along the lines
of~\cite{BreugelW14birthday,ChenBW12fossacs,Fu12icalp,TangB16concur,TangB17concur,TangB18cav}.
This direction, however, would require restricting our theory to
finite-state systems, which contrasts with our general motivation
coming from cryptographic security.  Indeed, in the analysis of
cryptographic protocols, security is usually to be proven against an
arbitrary adversary, hence also against infinite-state ones.  Hence,
model-checking of finite-state systems would not directly be
applicable in this setting.

\paragraph{Acknowledgements} 
Massimo Bartoletti is partially supported by 
Conv.\ Fondazione di Sardegna \& Atenei Sardi project
F75F21001220007 \emph{ASTRID}. 
Maurizio Murgia and Roberto Zunino are partially supported  PON \textit{Distributed Ledgers for Secure Open Communities}.
Maurizio Murgia is partially supported by MUR PON REACT EU DM 1062/21.

\bibliographystyle{alphaurl}
\bibliography{main}

\newpage
\appendix
\section{Proofs} \label{sec:proofs}

\begin{proofof}{Lemma}{lem:pctl:monotonicity}
  We simultaneously prove the whole statement by induction on the structure
  of the formulae $\logPhi$ and $\logPsi$.
  The cases $\logPhi=\labelL$ and $\logPhi=\sf true$ result
  in trivial equalities.
  For the case $\logPhi=\lnot\logPhii$ we need to prove
  \begin{align*}
    & \sem{\stepsN}{\distD}{+1}{\lnot\logPhii} \subseteq
    \sem{\stepsN}{\distDi}{+1}{\lnot\logPhii}
    \\
    & \sem{\stepsN}{\distDi}{-1}{\lnot\logPhii} \subseteq
    \sem{\stepsN}{\distD}{-1}{\lnot\logPhii}
    \\
    & \sem{\stepsN}{\distD}{-1}{\lnot\logPhii} \subseteq
    \sem{\stepsN}{\distD}{+1}{\lnot\logPhii}
  \end{align*}
  which is equivalent to
  \begin{align*}
    & \tsQ\setminus\sem{\stepsN}{\distD}{-1}{\logPhii} \subseteq
    \tsQ\setminus\sem{\stepsN}{\distDi}{-1}{\logPhii}
    \\
    & \tsQ\setminus\sem{\stepsN}{\distDi}{+1}{\logPhii} \subseteq
    \tsQ\setminus\sem{\stepsN}{\distD}{+1}{\logPhii}
    \\
    & \tsQ\setminus\sem{\stepsN}{\distD}{+1}{\logPhii} \subseteq
    \tsQ\setminus\sem{\stepsN}{\distD}{-1}{\logPhii}
  \end{align*}
  which, in turn, is equivalent to
  \begin{align*}
    & \sem{\stepsN}{\distDi}{-1}{\logPhii} \subseteq
    \sem{\stepsN}{\distD}{-1}{\logPhii}
    \\
    & \sem{\stepsN}{\distD}{+1}{\logPhii} \subseteq
    \sem{\stepsN}{\distDi}{+1}{\logPhii}
    \\
    & \sem{\stepsN}{\distD}{-1}{\logPhii} \subseteq
    \sem{\stepsN}{\distD}{+1}{\logPhii}
  \end{align*}
  which is the induction hypothesis.

  \noindent
  For the case $\logPhi=\logPhi_1 \land \logPhi_2$ we need to prove
  \begin{align*}
    & \sem{\stepsN}{\distD}{+1}{\logPhi_1\land\logPhi_2} \subseteq
    \sem{\stepsN}{\distDi}{+1}{\logPhi_1\land\logPhi_2}
    \\
    & \sem{\stepsN}{\distDi}{-1}{\logPhi_1\land\logPhi_2} \subseteq
    \sem{\stepsN}{\distD}{-1}{\logPhi_1\land\logPhi_2}
    \\
    & \sem{\stepsN}{\distD}{-1}{\logPhi_1\land\logPhi_2} \subseteq
    \sem{\stepsN}{\distD}{+1}{\logPhi_1\land\logPhi_2}
  \end{align*}
  which is equivalent to
  \begin{align*}
    & \sem{\stepsN}{\distD}{+1}{\logPhi_1}
    \cap \sem{\stepsN}{\distD}{+1}{\logPhi_2}
    \subseteq
    \sem{\stepsN}{\distDi}{+1}{\logPhi_1}
    \cap \sem{\stepsN}{\distDi}{+1}{\logPhi_2}
    \\
    & \sem{\stepsN}{\distDi}{-1}{\logPhi_1}
    \cap \sem{\stepsN}{\distDi}{-1}{\logPhi_2}
    \subseteq
    \sem{\stepsN}{\distD}{-1}{\logPhi_1}
    \cap \sem{\stepsN}{\distD}{-1}{\logPhi_2}
    \\
    & \sem{\stepsN}{\distD}{-1}{\logPhi_1}
    \cap \sem{\stepsN}{\distD}{-1}{\logPhi_2}
    \subseteq
    \sem{\stepsN}{\distD}{+1}{\logPhi_1}
    \cap \sem{\stepsN}{\distD}{+1}{\logPhi_2}
  \end{align*}
  which immediately follows from the induction hypothesis on
  $\logPhi_1$ and $\logPhi_2$.
  For the case $\logPhi=\logPr{\rhd \probP}{\logPsi}$ we need to prove
  \begin{align*}
    & \sem{\stepsN}{\distD}{+1}{\logPr{\rhd \probP}{\logPsi}} \subseteq
    \sem{\stepsN}{\distDi}{+1}{\logPr{\rhd \probP}{\logPsi}}
    \\
    & \sem{\stepsN}{\distDi}{-1}{\logPr{\rhd \probP}{\logPsi}} \subseteq
    \sem{\stepsN}{\distD}{-1}{\logPr{\rhd \probP}{\logPsi}}
    \\
    & \sem{\stepsN}{\distD}{-1}{\logPr{\rhd \probP}{\logPsi}} \subseteq
    \sem{\stepsN}{\distD}{+1}{\logPr{\rhd \probP}{\logPsi}}
  \end{align*}
  The first inclusion follows from
  \begin{align*}
    \sem{\stepsN}{\distD}{+1}{\logPr{\rhd \probP}{\logPsi}}
    & = \setcomp{\stateQ\in\tsQ}{
      \Pr(\trStart{\stateQ} \cap \sem{\stepsN}{\distD}{+1}{\logPsi})
      + \distD \rhd \probP }
    \\
    & \subseteq
    \setcomp{\stateQ\in\tsQ}{
      \Pr(\trStart{\stateQ} \cap \sem{\stepsN}{\distDi}{+1}{\logPsi})
      + \distDi \rhd \probP }
    \\
    & =
    \sem{\stepsN}{\distDi}{+1}{\logPr{\rhd \probP}{\logPsi}}
  \end{align*}
  where we exploited $\distD\leq\distDi$,
  the induction hypothesis
  $\sem{\stepsN}{\distD}{+1}{\logPsi} \subseteq
  \sem{\stepsN}{\distDi}{+1}{\logPsi}$, the monotonicity
  of $\Pr(-)$, and the fact that $\geq\circ\,\rhd \subseteq \rhd$.
  The second inclusion follows from an analogous argument:
  \begin{align*}
    \sem{\stepsN}{\distDi}{-1}{\logPr{\rhd \probP}{\logPsi}}
    & = 
    \setcomp{\stateQ\in\tsQ}{
      \Pr(\trStart{\stateQ} \cap \sem{\stepsN}{\distDi}{-1}{\logPsi})
      - \distDi \rhd \probP }
    \\
    & \subseteq
    \setcomp{\stateQ\in\tsQ}{
      \Pr(\trStart{\stateQ} \cap \sem{\stepsN}{\distD}{-1}{\logPsi})
      - \distD \rhd \probP }
    \\
    & =
    \sem{\stepsN}{\distD}{-1}{\logPr{\rhd \probP}{\logPsi}}
  \end{align*}
  where we exploited $-\distDi\leq-\distD$,
  the induction hypothesis
  $\sem{\stepsN}{\distDi}{-1}{\logPsi} \subseteq
  \sem{\stepsN}{\distD}{-1}{\logPsi}$, the monotonicity
  of $\Pr(-)$, and the fact that $\geq\circ\,\rhd \subseteq \rhd$.
  
  For $\logPsi = \logNext \logPhi$, we can observe that
  $\sem{\stepsN}{\distD}{\dirR}{\logNext \logPhi}
  = f(\sem{\stepsN}{\distD}{\dirR}{\logPhi})$
  where $f$ is a monotonic function mapping sets of states
  to sets of traces, which does not depend on $\distD,\dirR,\stepsN$.
  Hence, the thesis follows from the set inclusions about
  the semantics of $\logPhi$ in the induction hypothesis.

  Similarly, for $\logPsi = \logPhi_1 \logUntil \logPhi_2$, we can
  observe that
  \(
  \sem{\stepsN}{\distD}{\dirR}{\logPhi_1 \logUntil \logPhi_2} =
  g_n(\sem{\stepsN}{\distD}{\dirR}{\logPhi_1},
  \sem{\stepsN}{\distD}{\dirR}{\logPhi_2})
  \)
  where $g_n$ is a monotonic
  function mapping pairs of sets of states to sets of traces, which
  does not depend on $\distD,\dirR$ (but only on $\stepsN$).
  Hence, the thesis follows from the set inclusions about the
  semantics of $\logPhi_1$ and $\logPhi_2$ in the induction
  hypothesis.
  \qed
\end{proofof}

\begin{proofof}{Lemma}{lem:sim:monotonicity}
  The statement follows by induction on $\stepsN-\stepsNi$
  from the following properties:
  \begin{align}
    \label{eq:sim:monotonicity:1}
    & \distD \leq \distDi
    \;\land\;
    \stateP \crysim{\stepsN}{\distD} \stateQ
    \implies
    \stateP \crysim{\stepsN}{\distDi} \stateQ
    \\
    \label{eq:sim:monotonicity:2}
    & \stateP \crysim{\stepsN+1}{\distD} \stateQ
    \implies
    \stateP \crysim{\stepsN}{\distD} \stateQ
  \end{align}

  To prove \eqref{eq:sim:monotonicity:1} we proceed by induction on
  $\stepsN$.
  In the base case $\stepsN = 0$ the thesis trivially follows by the
  first case of Definition~\ref{def:param-bisim}.

  For the inductive case, we assume \eqref{eq:sim:monotonicity:1}
  holds for $\stepsN$, and prove it for $\stepsN+1$.
  Therefore, we assume $\stateP \crysim{\stepsN+1}{\distD} \stateQ$
  and prove $\stateP \crysim{\stepsN+1}{\distDi} \stateQ$.

  To prove the thesis, we must show that all the items of
  \autoref{def:param-bisim} hold.
  Item~\eqref{def:param-bisim:a} directly follows from the
  hypothesis.
  For item~\eqref{def:param-bisim:b} we have
  \[
  \tsPr{\stateP}{\statesQ}
  \leq \tsPr{\stateQ}{\cryset{\stepsN}{\distD}{\statesQ}} + \distD
  \leq \tsPr{\stateQ}{\cryset{\stepsN}{\distDi}{\statesQ}} + \distDi
  \]
  where the first inequality follows from the hypothesis
  $\stateP \crysim{\stepsN+1}{\distD} \stateQ$, while the second one
  follows from the induction hypothesis (which implies
  $\cryset{\stepsN}{\distD}{\statesQ} \subseteq
  \cryset{\stepsN}{\distDi}{\statesQ}$) and $\distD\leq\distDi$.
  Item~\eqref{def:param-bisim:c} is analogous.

  We now prove \eqref{eq:sim:monotonicity:2}, proceeding by induction
  on $\stepsN$.
  In the base case $\stepsN=0$, the thesis trivially follows by the
  first case of~\autoref{def:param-bisim}.
  For the inductive case, we assume the statement holds for $\stepsN$,
  and we prove it for $\stepsN+1$.
  Therefore, we assume $\stateP \crysim{\stepsN+2}{\distD} \stateQ$
  and prove $\stateP \crysim{\stepsN+1}{\distD} \stateQ$.

  To prove the thesis, we must show that all the items of
  \autoref{def:param-bisim} hold.
  Item~\eqref{def:param-bisim:a} directly follows from the hypothesis.
  For item~\eqref{def:param-bisim:b} of the thesis we have
  \[
  \tsPr{\stateP}{\statesQ}
  \leq \tsPr{\stateQ}{\cryset{\stepsN+1}{\distD}{\statesQ}} + \distD
  \leq \tsPr{\stateQ}{\cryset{\stepsN}{\distD}{\statesQ}} + \distD
  \]
  where the first inequality follows from the hypothesis
  $\stateP \crysim{\stepsN+2}{\distD} \stateQ$, while the second one
  follows from the induction hypothesis (which implies
  $\cryset{\stepsN+1}{\distD}{\statesQ} \subseteq
  \cryset{\stepsN}{\distD}{\statesQ}$).
  Item~\eqref{def:param-bisim:c} is analogous.
  \qed
\end{proofof}

\begin{samepage} 
\begin{applemma}\label{lem:leq-eps-implies-leq}
  Let $a,b \in \mathbb{R}$. If $\forall \epsilon > 0: a \leq b + \epsilon$ then $a \leq b$. 
\end{applemma}
\begin{proof}
  If $a>b$, taking $\epsilon=(a-b)/2$ contradicts the hypothesis.
\end{proof}
\end{samepage}

\begin{proofof}{Lemma}{lem:traces}
  By Lemma~\ref{lem:sim:monotonicity} we have that $\sP \CSim{m}{\distD} \sQ$.
  If $T$ is finite the thesis follows from Lemma~\ref{lem:finite-traces}. 
  If $T$ is infinite, it must be countable: this follows by the fact that
  Markov chains states are countable and the length of the traces in $T$ is finite. 
  So, let $\finTraceT_0 \finTraceT_1 \hdots$ be an enumeration of $T$. 
  By definition of infinite sum, we have that:
  \[
  \prob{T}{} = \lim_{k \to \infty} {\sum_{i = 0}^k \prob{\finTraceT_i}{}}
  \]
  By definition of limit of a sequence, we have that for all $\epsilon > 0$ there exists $v \in \mathbb{N}$ such that for all $k > v$:
  \[
  \abs{\prob{T}{} - \sum_{i = 0}^k \prob{\finTraceT_i}{}} < \epsilon
  \]
  Since $\prob{\finTraceT_i}{} \geq 0$ for all $i$, we can drop the absolute value and we get:
  \begin{equation}\label{lem:traces:eq1}
    \prob{T}{} - \sum_{i = 0}^k \prob{\finTraceT_i}{} < \epsilon
  \end{equation}
  By Lemma~\ref{lem:leq-eps-implies-leq} it suffice to show $\prob{T}{} \leq \prob{\TR{m}{\distD,\sQ}{T}}{} + \distD m + \epsilon$ for all $\epsilon > 0$, or equivalently:
  \[
  \prob{T}{} - \epsilon \leq \prob{\TR{m}{\distD,\sQ}{T}}{} + \distD m
  \]
  So, let $\epsilon > 0$ and let $k$ be such that Lemma~\ref{lem:traces:eq1} holds.
  Then we have:
  \[
  \prob{T}{} - \epsilon < \sum_{i = 0}^k \prob{\finTraceT_i}{}
  \]
  Let $T' = \setcomp{\finTraceT_i}{i \leq k}$.
  Since $\sum_{i = 0}^k \prob{\finTraceT_i}{} = \prob{T'}{}$ and $T'$ is finite, by Lemma~\ref{lem:finite-traces} we have:
  \[
  \sum_{i = 0}^k \prob{\finTraceT_i}{} \leq \prob{\TR{m}{\distD,\sQ}{T'}}{} + \distD m
  \]
  Since $\TR{m}{\distD,\sQ}{T'} \subseteq \TR{m}{\distD,\sQ}{T}$ we have that:
  \[
  \prob{\TR{m}{\distD,\sQ}{T'}}{} + \distD m \leq \prob{\TR{m}{\distD,\sQ}{T}}{} + \distD m
  \]
  Summing up, we have that
  $\prob{T}{} - \epsilon \leq \prob{\TR{m}{\distD,\sQ}{T}}{} + \distD m$
  for all $\epsilon > 0$.
  By Lemma~\ref{lem:leq-eps-implies-leq} it follows that
  $\prob{T}{} \leq \prob{\TR{m}{\distD,\sQ}{T}}{} + \distD m$ as required.
  \qed
\end{proofof}

\begin{proofof}{Lemma}{lem:matching}
  Without loss of generality, we prove the statement under the following additional assumptions:
  \begin{align}
    \label{eq:matching-aux2}
    & \forall b \in B: f_B(b) > 0
    \\
    \label{eq:matching-aux1}
    & \forall b \in B: \setcomp{a \in A}{b \in g(a)} \neq \emptyset  \qquad \text{and}
    \\
    \nonumber
    & \qquad \forall b_1,b_2 \in B: \setcomp{a \in A}{b_1 \in g(a)} = \setcomp{a \in A}{b_2 \in g(a)} \implies b_1 = b_2
  \end{align}
  If $B$ does not satisfy \autoref{eq:matching-aux2}, just remove from $B$ the elements $b$ such that $f_B(b) = 0$ adjust $g$ accordingly, and set $h(a,b) = 0$.
  \autoref{eq:matching-assumption} still holds since we removed only elements whose value is zero. 
  If $B$ does not satisfy \autoref{eq:matching-aux1}, it can be transformed to a set that does. To see why, 
  let $\equiv \subseteq B \times B$ be defined as: 
  \[
  b \equiv b'
  \text{ iff } 
  \setcomp{a \in A}{b \in g(a)} = \setcomp{a \in A}{b' \in g(a)}
  \] 
  Let $\hat{B}$ be the set of equivalence classes \wrt $\equiv$.
  For an equivalence class $[b]$, define:
  \[
  f_{\hat{B}}([b]) = \sum_{b' \in [b]}{f_B(b')}
  \qquad\qquad
  g'(a) = \setcomp{[b]}{b \in g(a)}
  \]
  It is easy to verify that~\eqref{eq:matching-aux1} is satisfied.
  Notice that $\sum_{[b] \in g'(a)} f_{\hat{B}}([b])$ converges, since: 
  \[
  \sum_{[b] \in g'(a)} f_{\hat{B}}([b]) = \sum_{[b] \in g'(a)} \sum_{b' \in [b]}f_{B}(b) = \sum_{b \in g(a)}f_{B}(b)
  \]
  We now show that $A,\hat{B}$ and $g'$ satisfy \autoref{eq:matching-assumption}. We have that, for all $b \in B$, 
  $f_B(b) \leq f_{\hat{B}}([b])$ and $b \in g(a) \implies [b] \in g'(a)$.
  Therefore, for all $A' \subseteq A$:
  \[
  \sum_{a \in A'} f_A(a) \leq \sum_{b \in \bigcup_{a \in A'} g(a)} f_B(b) \leq \sum_{[b] \in \bigcup_{a \in A'} g'(a)} f_{\hat{B}}([b])
  \]
  From a function $h'$ satisfying
  \autoref{eq:matching-thesis:1} and~\autoref{eq:matching-thesis:2}
  for $A, \hat{B}$
  and $g'$ we can easily obtain a function $h$ for $A, B$ and $g$:
  \eg, set $h(a,b) = h'(a,[b])\frac{f_B(b)}{f_{\hat{B}}([b])}$.
  Notice that $f_{\hat{B}}([b]) > 0$ by \autoref{eq:matching-aux2}, and
  that if $B$ satisfies \autoref{eq:matching-aux1} it then holds that
  $\card{B} < 2^{\card{A}}$, and so $B$ is finite.  That said, we show
  that the thesis holds by reducing to the max-flow problem~\cite{MinCut}.  
  Assume w.l.o.g.\ that $A$ and $B$ are disjoint. 
  Let $N = (V,E)$ be a directed graph, where
  $V = A \cup B \cup \setenum{s,t}$ with $s,t \not\in A \cup B$ and:
  \[
  E = \setcomp{(s,b)}{b \in B} \cup \setcomp{(b,a)}{a \in A, b \in g(a)} \cup \setcomp{(a,t)}{a \in A}
  \]
  Define edge capacity $w: E \rightarrow \NNR \cup \setenum{\infty}$ as follows:
  \[
  w(s,b) = f_B(b)
  \qquad
  w(b,a) = \infty
  \qquad
  w(a,t) = f_A(a)
  \]
  Consider the cut $C = \setcomp{(a,t)}{a \in A}$ associated with partition $(V \setminus \setenum{t},\setenum{t})$. Such cut has capacity
  $\sum_{a \in A} f_A(a)$ and we argue it is minimum. Take a cut $C'$ of the network. First notice that if $C'$ contains edges of the form
  $(b,a)$ its capacity would be infinite.
  We can therefore consider only cuts whose elements
  are of the form $(s,b)$ or $(a,t)$, and thus for all $a \in A$ we have that $a$ and the elements of $g(a)$ are in the same partition.
  In other words, $s$ partition is of the form $A' \cup \bigcup_{a \in A'} g(a) \cup \setenum{s}$, $t$ partition is of the form
  $A \setminus A' \cup \bigcup_{a \in (A \setminus A')} g(a) \cup \setenum{t}$, where $A' \subseteq A$.
  So capacity of $C'$ is $\sum_{a \in A'} f_A(a) + \sum_{b \in g(A \setminus A')} f_B(b)$. Now, capacity of $C$ is 
  $\sum_{a \in A'} f_A(a) + \sum_{a \in (A \setminus A')} f_A(a)$. Since $\sum_{a \in (A \setminus A')} f_A(a) \leq \sum_{b \in g(A \setminus A')} f_B(b)$
  by assumption \autoref{eq:matching-assumption}, we have that capacity of $C$ is minimal.
  By the min-cut max-flow theorem \cite{MinCut},
  we have that the max flow of the network has capacity $\sum_{a \in A} f_A(a)$.

  Let $\flow: E \rightarrow \NNR$ be the a flow associated to such cut.
  Consequently, we have that $\flow(a,t) = f_A(a)$ for all $a \in A$.
  Define: 
  \[
  h(a,b) =
  \begin{cases}
    \frac{\flow(b,a)}{f_B(b)} & \text{ if } b \in g(a)\\
    0 & \text{ otherwise}
  \end{cases} 
  \]
  We have to show that $h$ satisfies
  \autoref{eq:matching-thesis:1} and~\autoref{eq:matching-thesis:2}.
  Let $A' \subseteq A$. We have that:
  \begin{align*}
    \sum_{a \in A'} \sum_{b \in g(a)} h(a,b) f_B(b) 
    & = \sum_{a \in A'} \sum_{b \in g(a)} \frac{\flow(b,a)}{f_B(b)} f_B(b) 
    \\
    & = \sum_{a \in A'} \sum_{b \in g(a)} \flow(b,a)
  \end{align*}
  By the conservation of flow constraint, we have that: 
  \begin{align*}
    \sum_{a \in A'} \sum_{b \in g(a)} \flow(b,a)
    & =
    \sum_{a \in A'} \flow(a,t)
    \\
    & =
    \sum_{a \in A'} f_A(a)
  \end{align*}
  So summing up we have that:
  \[
  \sum_{a \in A'} \sum_{b \in g(a)} h(a,b) f_B(b) = \sum_{a \in A'} f_A(a)
  \] 
  For the remaining part, let $b \in B$. We have that:
  \begin{align*}
    \sum_{a \in A} h(a,b) 
    & = \sum_{a \in \setcomp{a'}{b \in g(a')}} h(a,b)
    \\
    & = \sum_{a \in \setcomp{a'}{b \in g(a')}} \frac{\flow(b,a)}{f_B(b)} 
    \\
    & = \frac{1}{f_B(b)} {\sum_{a \in \setcomp{a'}{b \in g(a')}} \flow(b,a)}
    \\
    & \leq\; \frac{f_B(b)}{f_B(b)}
    \\
    & = \; 1
    \tag*{\qed}
  \end{align*}
\end{proofof}

\begin{proofof}{Lemma}{lem:finite-traces}
  By induction on $n$.
  The base case ($n = 1$) is trivial as $T = \{\sP\}$ and $\TR{n}{\distD,\sQ}{T} = \{\sQ\}$, 
  or $T = \emptyset$ and $\TR{n}{\distD,\sQ}{T} = \emptyset$.
  Therefore,
  $\prob{T}{} = \prob{\TR{n}{\distD,\sQ}{T}}{} = \card{T}$.
  For the inductive case, first notice that:
  \[
  \prob{T}{}\;\;= \;\;
  \sum_{\finTraceT \in T} \prob{\sP}{\finTraceT(1)}\prob{\finTraceT(1 .. n - 1)}{}
  \]
  Referring to Lemma~\ref{lem:matching},
  let $A = \setcomp{\finTraceT(1)}{\finTraceT \in T}$, 
  $B = \setcomp{\sQi}{\sPi \CSim{n - 1}{\distD} \sQi \text{ for some } \sPi \in A} \cup \setenum{D}$,
  where $D$ is a special element not occurring in $A \cup B$.
  Let $f_A(\sPi) = \prob{\sP}{\sPi}$, $f_B(\sQi) = \prob{\sQ}{\sQi}$ and
  $f_B(D) = \distD$.
  Finally, let $g(\sPi) = \;\R{n-1}{\distD}{\sPi} \cup \setenum{D}$. 

  By~\autoref{def:param-bisim}, 
  we have that $A, B, f_A, f_B$ and $g$ satisfy
  \autoref{eq:matching-assumption} of \autoref{lem:matching}.
  Indeed, for all $A' \subseteq A$, we have that: 
  \[
  \sum_{a \in A'} f_A(a) = \prob{\sP}{A'} \leq \prob{\sQ}{\cryset{n - 1}{\distD}{A'}} + \distD = \sum_{b \in \bigcup_{a \in A'} g(a)} f_B(b)
  \]
  We can then conclude that there exist $h$ such that,
  for all $A' \subseteq A$:
  \[
  \prob{\sP}{A'} = \sum_{\sPi \in A'} \Big( h(\sPi,D) \distD \;\;+
  \sum_{\sQi \in \R{n-1}{\distD}{\sPi}} h(\sPi,\sQi)\prob{\sQ}{\sQi}\Big)
  \]
  Let $T_{\sSetP} = \setcomp{\finTraceT(1..n - 1)}{\finTraceT \in T \,\land\, \finTraceT(1) \in \sSetP}$ 
  where $\sSetP \subseteq A$. 
  We simply write $T_{\sPi}$ if $\sSetP = \{\sPi\}$. 
  So, we have that:
  \begin{align*}
    \prob{T}{} 
    & \;\;=\;\;\sum_{\finTraceT \in T} \prob{\sP}{\finTraceT(1)} \prob{\finTraceT(1.. n - 1)}{}
    \\
    & \;\;= \;\; \sum_{\sPi \in A} \prob{\sP}{\sPi}\prob{T_{\sPi}}{}
    \\
    & \;\;= \;\; \sum_{\sPi \in A} \prob{T_{\sPi}}{}\Big(h(\sPi,D) \distD \;\;+ \sum_{\sQi \in \R{n-1}{\distD}{\sPi}} h(\sPi,\sQi)\prob{\sQ}{\sQi}\Big)
    \\
    & \;\;\leq \;\; \distD + \sum_{\sPi \in A} \prob{T_{\sPi}}{}\sum_{\sQi \in \R{n-1}{\distD}{\sPi}} h(\sPi,\sQi)\prob{\sQ}{\sQi}
    \\
    & \;\;= \;\; \distD + \sum_{\sPi \in A} \sum_{\sQi \in \R{n-1}{\distD}{\sPi}} h(\sPi,\sQi)\prob{\sQ}{\sQi}\prob{T_{\sPi}}{}
    \\
    & \;\;\leq \;\; \distD + \sum_{\sPi \in A} \sum_{\sQi \in \R{n-1}{\distD}{\sPi}} h(\sPi,\sQi)\prob{\sQ}{\sQi}\Big(\prob{\TR{n-1}{\distD,\sQi}{T_{\sPi}}}{} + \distD(n - 1)\Big)
    \\
    & \;\;= \;\; \distD + s_1 + s_2
  \end{align*}
  where:
  \begin{align*}
    s_1 & = \sum_{\sPi \in A} \sum_{\sQi \in \R{n-1}{\distD}{\sPi}} h(\sPi,\sQi)\prob{\sQ}{\sQi} \distD (n - 1)
    \\
    s_2 & = \sum_{\sPi \in A} \sum_{\sQi \in \R{n-1}{\distD}{\sPi}} h(\sPi,\sQi)\prob{\sQ}{\sQi}\prob{\TR{n-1}{\distD,\sQi}{T_{\sPi}}}{}
  \end{align*}
  Now:
  \begin{align*}
    s_1
    & = \distD (n - 1)\sum_{\sPi \in A} \sum_{\sQi \in \R{n-1}{\distD}{\sPi}} h(\sPi,\sQi)\prob{\sQ}{\sQi}
    \\
    & \leq \distD (n - 1)\prob{\sP}{A}
    \\
    & \leq \distD (n - 1)
  \end{align*}
  Therefore $\distD + s_1 \leq \distD n$.
  It remains to show that $s_2 \leq \prob{\TR{n}{\distD,\sQ}{T}}{}$. 
  First notice that $s_2$ can be rewritten as follows by a simple reordering of terms:
  \[
  s_2 = \sum_{\sQi \in \R{n - 1}{\distD}{A}} \sum_{\sPi \in A \cap \R{n - 1}{\distD}{\sQi}} h(\sPi,\sQi)\prob{\sQ}{\sQi}\prob{\TR{n-1}{\distD,\sQi}{T_{\sPi}}}{}
  \]
  So:
  \begin{align*}
    s_2
    & = \sum_{\sQi \in \R{n - 1}{\distD}{A}}
    \quad \sum_{\sPi \in A \cap \R{n - 1}{\distD}{\sQi}}
    h(\sPi,\sQi)\prob{\sQ}{\sQi}\prob{\TR{n-1}{\distD,\sQi}{T_{\sPi}}}{}
    \\
    & \leq \sum_{\sQi \in \R{n - 1}{\distD}{A}}
    \quad \sum_{\sPi \in A \cap \R{n - 1}{\distD}{\sQi}}
    h(\sPi,\sQi)\prob{\sQ}{\sQi}
    \prob{\TR{n-1}{\distD,\sQi}{T_{A \cap \R{n - 1}{\distD}{\sQi}}}}{}
    \\
    & \leq \sum_{\sQi \in \R{n - 1}{\distD}{A}} \prob{\sQ}{\sQi}\prob{\TR{n-1}{\distD,\sQi}{T_{A \cap \R{n - 1}{\distD}{\sQi}}}}{}
    \sum_{\sPi \in A \cap \R{n - 1}{\distD}{\sQi}} h(\sPi,\sQi)
    \\
    & \leq \sum_{\sQi \in \R{n - 1}{\distD}{A}} \prob{\sQ}{\sQi}\prob{\TR{n-1}{\distD,\sQi}{T_{A \cap \R{n - 1}{\distD}{\sQi}}}}{}
    \\
    & = \prob{\TR{n}{\distD,\sQ}{T}}{}
  \end{align*}
  The last equality follows by partitioning $\TR{n}{\distD,\sQ}{\tracesT}$
  according to the second state of each trace $\sQi$.  The set of all
  such second states is the set of those bisimilar to (some state of)
  $A$, namely $\R{n - 1}{\distD}{A}$.
  Given any such $\sQi$, the probability of its partition is
  $\prob{\sQ}{\sQi}\prob{U_{\sQi}}{}$ where $U_{\sQi}$ is the set of
  the \emph{tails} of $\TR{n}{\distD,\sQ}{\tracesT}$ starting from $\sQi$.
  Since this set is defined taking pointwise bisimilar traces, we can
  equivalently express $U_{\sQi}$ by first taking the tails of
  $\tracesT$ (i.e., $T_A$), and then considering the bisimilar traces:
  in other words, we have $U_{\sQi} = \TR{n-1}{\distD,\sQi}{T_{A}}$.
  Note that the states in $A$ which are not bisimilar to $\sQi$ do not
  contribute to $\TR{n-1}{\distD,\sQi}{T_{A}}$ in any way, so we can also
  write the desired
  $U_{\sQi} = \TR{n-1}{\distD,\sQi}{T_{A \cap \R{n - 1}{\distD}{\sQi}}}$.
  \qed
\end{proofof}

\begin{applemma}\label{lem:logic-finite-traces-next}
  Let $T = \setcomp{\traceT}{\traceT(0) = \sP \,\land\, \traceT \sat{\distD}{n}{r} \lNext \sFormula}$ for some $\sP,\sFormula$, and let $m \geq 2$. Then:
  \[
  \prob{T}{} = \prob{\setcomp{\finTraceT}{\card{\traceT} = m \,\land\, \finTraceT(0) = \sP \,\land\, \finTraceT \sat{\distD}{n}{r} \lNext \sFormula}}{}
  \]
\end{applemma}
\begin{proof}
  Trivial.
\end{proof}

\begin{applemma}\label{lem:logic-finite-traces}
  Let $T = \setcomp{\traceT}{\traceT(0) = \sP \,\land\, \traceT \sat{\distD}{n}{r} \sFormula[1] \lUntil \sFormula[2]}$ for some 
  $\sP,\sFormula[1], \sFormula[2]$, and let $m \geq n + 1$. Then:
  \[
  \prob{T}{}
  =
  \prob{\setcomp{\finTraceT}{\card{\finTraceT} = m \,\land\, \finTraceT(0) = \sP \,\land\, \finTraceT \sat{\distD}{n}{r} \sFormula[1] \lUntil \sFormula[2]}}{}
  \]
\end{applemma}
\begin{proof}
  (Sketch)
  Let $\tilde{T} = \setcomp{\finTraceT}{\card{\finTraceT} = m \,\land\, \finTraceT(0) = \sP \,\land\, \finTraceT \sat{\distD}{n}{r} \sFormula[1] \lUntil \sFormula[2]}$.
  The thesis follows from the fact that $T = \bigcup_{\finTraceT \in \tilde{T}}{\cyl{\finTraceT}}$.
\end{proof}

\noindent
For notational convenience, hereafter we will often write
$\stateQ \sat{\distD}{\stepsN}{\dirR} \logPhi$ instead of
$\stateQ \in \sem{\stepsN}{\distD}{\dirR}{\logPhi}$.

\begin{applemma}\label{lem:bisimi-implies-prop-preserv}
  Let $k$ and $n$ be, respectively, the maximum nesting level of $\lUntil$ and of $\lNext$ in $\sFormula$, and let
  $\sP \CSim{mk + n + 1}{\distD_1} \sQ$. Then:
  \begin{enumerate}
  \item \label{lem:bisimi-implies-prop-preserv:item1}
    $\sP \sat{\distD_2}{m}{+1} \sFormula \implies \sQ \sat{\distD_2 + \distD_1(mk + n + 1)}{m}{+1} \sFormula$
  \item \label{lem:bisimi-implies-prop-preserv:item2}
    $\sP \not\sat{\distD_2}{m}{-1} \sFormula \implies \sQ \not\sat{\distD_2 + \distD_1(mk + n + 1)}{m}{-1} \sFormula$
  \end{enumerate}
\end{applemma}
\begin{proof}
  By induction on $\sFormula$. The cases $\lTrue$ and $\atomA$ are trivial.
  \begin{itemize}

  \item $\lNot \sFormulai$.
    We only show \autoref{lem:bisimi-implies-prop-preserv:item1} as the other item is similar.
    So, suppose $\sP \sat{\distD_2}{m}{+1} \lNot \sFormulai$.
    Then, $\sP \not\sat{\distD_2}{m}{-1} \sFormula$.
    By the induction hypothesis we have that 
    $\sQ \not\sat{\distD_2 + \distD_1(mk + n + 1)}{m}{-1} \sFormula$, and
    hence $\sQ \sat{\distD_2 + \distD_1(mk + n + 1)}{m}{+1} \lNot \sFormulai$ as required.

  \item $\sFormula[1] \,\lAnd\, \sFormula[2]$.
    We only show \autoref{lem:bisimi-implies-prop-preserv:item1} as the other item is similar.
    So, suppose $\sP \sat{\distD_2}{m}{+1} \sFormula[1] \,\lAnd\, \sFormula[2]$. Then $\sP \sat{\distD_2}{m}{+1} \sFormula[1]$ and $\sP \sat{\distD_2}{m}{+1} \sFormula[2]$.
    By the induction hypothesis $\sQ \sat{\distD_2 + \distD_1(mk + n + 1)}{m}{+1} \sFormula[1]$ and 
    $\sQ \sat{\distD_2 + \distD_1(mk + n + 1)}{m}{+1} \sFormula[2]$.
    Therefore
    $\sQ \sat{\distD_2 + \distD_1(mk + n + 1)}{m}{+1} \sFormula[1] \,\lAnd\, \sFormula[2]$ as required.

  \item $\logPr{\rhd \probP}{\pFormula}$.
    For \autoref{lem:bisimi-implies-prop-preserv:item1},
    suppose that $\sP \sat{\distD_2}{m}{+1} \logPr{\rhd \probP}{\pFormula}$.
    We only deal with the case $\rhd = \,\geq$,
    since the case $\rhd = \, >$ is analogous.
    Let: 
    \[
    T = \setcomp{\finTraceT}{\card{\finTraceT} = mk + n + 1 \,\land\, \finTraceT(0) = \sP \,\land\, \finTraceT \sat{\distD_2}{m}{+1} \pFormula}{}
    \]
    We start by proving that:
    \begin{equation}\label{lem:bisimi-implies-prop-preserv:eq1}
      \forall \finTraceU \in \TR{mk + n + 1}{\distD_1,\sQ}{T}
      \;\; : \;\;
      \finTraceU \sat{\distD_2 + \distD_1(mk + n + 1)}{m}{+1} \pFormula
    \end{equation}
    Let $\finTraceU \in \TR{m k + n + 1}{\distD_1,\sQ}{T}$. Then, there is $\finTraceT \in T$ such that,
    for all $0 \leq i < mk + n + 1$: 
    \[\finTraceT(i)
    \CSim{mk + n + 1-i}{\distD_1} \finTraceU(i)\] 
    We proceed by cases on $\pFormula$.
    \begin{itemize}
      
    \item
      $\sFormula[1] \lUntil \sFormula[2]$.
      First notice that $mk + n + 1 \geq m + 1$, and hence by \autoref{lem:logic-finite-traces} we have that: 
      \[\prob{T}{} = \prob{\setcomp{\traceT}{\traceT(0) = \sP \,\land\, \traceT \sat{\distD_2}{m}{+1} \sFormula[1] \lUntil \sFormula[2]}}{}\]
      We then  have $\prob{T}{} + \distD_2 \geq \probP$.
      Since $\finTraceT \sat{\distD_2}{m}{+1} \sFormula[1] \lUntil \sFormula[2]$, we have that: 
      \[
      \exists i \leq m: 
      \finTraceT(i) \sat{\distD_2}{m}{+1} \sFormula[2] \,\land\, \forall j < i: \finTraceT(j) \sat{\distD_2}{m}{+1} \sFormula[1]
      \]
      Let $n'$ be the maximum nesting level of $\lNext$ in $\sFormula[2]$.
      We know that: 
      \[
      \finTraceT(i) \CSim{mk + n + 1-i}{\distD_1} \finTraceU(i) \,\land\, mk + n + 1 - i > m(k - 1) + n' + 1
      \] 
      Then, by Lemma~\ref{lem:sim:monotonicity}
      (monotonicity of $\CSim{}{}$), we have that:
      \[
      \finTraceT(i) \CSim{m(k - 1) + n' + 1}{\distD_1} \finTraceU(i)
      \] 
      Then, by the induction hypothesis, we have that: 
      \[
      \finTraceU(i) \sat{\distD_2 + \distD_1(m(k - 1) + n' + 1)}{m}{+1} \sFormula[2]
      \]
      By Lemma~\ref{lem:pctl:monotonicity} (monotonicity of $\sat{}{}{}$)
      it follows that: 
      \[
      \finTraceU(i) \sat{\distD_2 + \distD_1(mk + n + 1)}{m}{+1} \sFormula[2]
      \] 
      With a similar argument we can conclude that, for all $j < i$:
      \[
      \finTraceU(j) \sat{\distD_2 + \distD_1(mk + n + 1)}{m}{+1} \sFormula[1]
      \]
      Hence \autoref{lem:bisimi-implies-prop-preserv:eq1} holds.
      
    \item
      $\lNext \sFormula[1]$.
      First notice that $mk + n + 1 \geq 2$, and hence by \autoref{lem:logic-finite-traces-next} we have that: 
      \[
      \prob{T}{} = \prob{\setcomp{\traceT}{\traceT(0) = \sP \,\land\, \traceT \sat{\distD_2}{m}{+1} \lNext \sFormula[1]}}{}
      \]
      Then, $\prob{T}{} + \distD_2 \geq \probP$.
      Since $\finTraceT \sat{\distD_2}{m}{+1} \lNext \sFormula[1]$, we have that
      \( \finTraceT(1) \sat{\distD_2}{m}{+1} \sFormula[1] \).
      We know that
      \(
      \finTraceU(1) \CSim{mk + n}{\distD_1} \finTraceT(1)
      \). 
      By the induction hypothesis,
      \(
      \finTraceU(1) \sat{\distD_2 + \distD_1(mk + n)}{m}{+1} \sFormula[1]
      \).
      By Lemma~\ref{lem:pctl:monotonicity} (monotonicity of $\sat{}{}{}$)
      it follows that: 
      \( \finTraceU(i) \sat{\distD_2 + \distD_1(mk + n + 1)}{m}{+1} \sFormula[1] \).
      Hence, \eqref{lem:bisimi-implies-prop-preserv:eq1} holds.
    \end{itemize} 
    
    \medskip
    Back to the main statement, we have that, by Lemma~\ref{lem:traces}:
    \[
    \prob{\TR{mk + n + 1}{\distD_1,\sQ}{T}}{} + \distD_2 + \distD_1 (mk + n + 1) 
    \geq 
    \prob{T}{} + \distD_2
    \]
    So, summing up:
    \begin{align*}
      & \prob{\setcomp{\traceT}{\traceT(0) = \sQ \;\land\; \traceT\sat{\distD_2 + \distD_1(mk + n + 1)}{m}{+1} \pFormula}}{} + \distD_2 + \distD_1 (mk + n + 1) 
      \\
      = \; & \prob{\setcomp{\finTraceT}{\card{\finTraceT} = mk + n + 1 \;\land\; 
          \finTraceT(0) = \sQ \;\land\; \finTraceT\sat{\distD_2 + \distD_1(mk + n + 1)}{m}{+1} \pFormula}}{}
      \\
      & \;\;\;\; + \distD_2 + \distD_1 (mk + n + 1) \\
      \geq \; &
      \prob{\TR{mk + n + 1}{\distD_1,\sQ}{T}}{} + \distD_2 + \distD_1 (mk + n + 1) 
      \\
      \geq \; & \prob{T}{} + \distD_2 
      \\
      \geq \; & \probP
    \end{align*}
    Therefore,
    $\sQ \sat{\distD_2 + \distD_1(mk + n)}{m}{+1} \logPr{\geq \probP}{\pFormula}$.

    \medskip
    For \autoref{lem:bisimi-implies-prop-preserv:item2}, suppose that
    $\sP \not\sat{\distD_2}{m}{-1} \logPr{\geq \probP}{\pFormula}$.
    Then:
    \[
    \prob{\setcomp{\traceT}{\traceT(0) = \sP \,\land\, \traceT \sat{\distD_2}{m}{-1} \pFormula}{}}{} - \distD_2 < \probP
    \]
    From the above, by a case analysis on $\pFormula$, and exploiting
    \autoref{lem:logic-finite-traces} and \autoref{lem:logic-finite-traces-next}, 
    we conclude that $\prob{T}{} - \distD_2 < \probP$, where: 
    \[
    T = \setcomp{\finTraceT}{\card{\finTraceT} = mk + n + 1 \,\land\, \traceT(0) = \sP \,\land\, \traceT \sat{\distD_2}{m}{-1} 
      \pFormula}
    \]
    Let: 
    \[\bar T = \setcomp{\finTraceT}{\card{\finTraceT} = mk + n + 1 \,\land\, t(0) = \sP \,\land\, \finTraceT \not\sat{\distD_2}{m}{-1} 
      \pFormula}{}\]
    We have that $1 - \prob{\bar T}{} = \prob{T}{}$. 
    We start by proving that: 
    \[
    \forall \finTraceU \in \TR{mk + n + 1}{\distD_1,\sQ}{\bar T}
    \;\; : \;\;
    \finTraceU \not\sat{\distD_2 + \distD_1(mk + n + 1)}{m}{-1} \pFormula
    \]
    Let $\finTraceU \in \TR{mk + n + 1}{\distD_1,\sQ}{\bar T}$.
    Then, there exist $\finTraceT \in \bar{T}$ such that, 
    for all $0 \leq i < mk + n + 1$: 
    \[\finTraceT(i) \CSim{mk + n-i}{\distD_1} \finTraceU(i)\]
    We proceed by cases on $\pFormula$.
    \begin{itemize}

    \item $\sFormula[1] \lUntil \sFormula[2]$.
      Since $\finTraceT \not\sat{\distD_2}{m}{-1} \sFormula[1] \lUntil \sFormula[2]$, we have that:
      \[
      \forall i \leq m: \finTraceT(i) \not\sat{\distD_1}{m}{-1} \sFormula[2] \lor \exists j < i: \finTraceT(j) \not\sat{\distD_2}{m}{-1} \sFormula[1]
      \]
      Take $i \leq m$.
      Let $n'$ be the maximum nesting level of $\lNext$ in $\sFormula[2]$. 
      If $\finTraceT(i) \not\sat{\distD_1}{m}{-1} \sFormula[2]$, since
      \[
      \finTraceT(i) \CSim{mk + n + 1 - i}{\distD_1} \finTraceU(i) \,\land\, mk + n + 1 - i > m(k - 1) + n' + 1
      \] 
      by Lemma~\ref{lem:sim:monotonicity} (monotonicity of $\CSim{}{}$)
      we have that: 
      \[
      \finTraceT(i) \CSim{m(k - 1) + n' + 1}{\distD_1} \finTraceU(i)
      \] 
      By the induction hypothesis we have that: 
      \[
      \finTraceU(i) \not\sat{\distD_2 + \distD_1(m(k - 1) + n' + 1)}{m}{-1} \sFormula[2]
      \]
      By Lemma~\ref{lem:pctl:monotonicity} (monotonicity of $\sat{}{}{}$) it follows: 
      \[
      \finTraceU(i) \not\sat{\distD_2 + \distD_1(mk + n + 1)}{m}{-1} \sFormula[2]
      \]
      If $\finTraceT(j) \not\sat{\distD_1}{m}{-1} \sFormula[1]$ for
      some $j < i$, with a similar argument we can conclude that: 
      \[
      \finTraceU(j) \not\sat{\distD_2 + \distD_1(m(k - 1) + n + 1)}{m}{-1} \sFormula[1]
      \]

    \item $\lNext \sFormula[1]$.
      Since $\finTraceT \not\sat{\distD_2}{m}{-1} \lNext \sFormula[1]$,
      we have that:
      \(
      \finTraceT(1) \not\sat{\distD_2}{m}{-1} \sFormula[1]
      \).
      Since $\finTraceT(1) \CSim{mk + n}{\distD_1} \finTraceU(1)$, 
      by the induction hypothesis we have
      \( \finTraceU(i) \not\sat{\distD_2 + \distD_1(mk + n)}{m}{-1} \sFormula[1] \).
      By Lemma~\ref{lem:pctl:monotonicity} it follows that:
      \[
      \finTraceU(i) \not\sat{\distD_2 + \distD_1(mk + n + 1)}{m}{-1} \sFormula[1]
      \]

    \end{itemize}

    \medskip
    Back to the main statement, by Lemma~\ref{lem:traces} we have that:
    \[\prob{\bar T}{} \leq \prob{\TR{mk + n + 1}{\distD_1,\sQ}{\bar T}}{} + \distD_1(mk + n + 1)\] 
    Summing up, we have that:
    \begin{align*}
      \hspace{-12pt}
      & \prob{\setcomp{\traceT}{\traceT(0) = \sQ \,\land\, \traceT\sat{\distD_2 + \distD_1(mk + n + 1)}{m}{-1} \pFormula}}{} - \distD_2 - \distD_1 (mk + n + 1) 
      \\ 
      & = \prob{\setcomp{\finTraceT}{\card{\finTraceT} = \card{\finTraceT} = mk + n + 1 \,\land\, \finTraceT(0) = \sQ \,\land\, \finTraceT\sat{\distD_2 + \distD_1(mk + n + 1)}{m}{-1} \pFormula}}{} 
      \\ 
      & \hspace{12pt} - \distD_2 - \distD_1 (mk + n + 1) 
      \\
      & = 1 - \prob{\setcomp{\finTraceT}{\card{\finTraceT} = mk + n + 1 \,\land\, \finTraceT(0) = \sQ \,\land\, \finTraceT\not\sat{\distD_2 + \distD_1(mk + n + 1)}{m}{-1} \pFormula}}{} 
      \\
      & \hspace{12pt} - \distD_2 - \distD_1 (mk + n + 1) 
      \\
      & \leq
      1 - \prob{\TR{mk + n}{\distD_1,\sQ}{\bar T}}{} - \distD_2 - \distD_1 (mk + n + 1) 
      \\
      & \leq
      1 - \prob{\bar T}{} - \distD_2 
      \\
      & = \prob{T}{} - \distD_2
      <
      \probP
    \end{align*}
    Therefore,
    $\sQ \not\sat{\distD_2 + \distD_1(mk + n)}{m}{-1} \logPr{\geq \probP}{\pFormula}$.
    \qedhere
  \end{itemize}
\end{proof}

\begin{proofof}{Theorem}{th:soundness}
  Immediate consequence of~\autoref{lem:bisimi-implies-prop-preserv}.\qed

  \ 
\end{proofof}

\end{document}